\documentclass[intlimits,twoside,a4paper]{article}
\usepackage{amsmath,amssymb}
\usepackage{graphicx}
\usepackage{wrapfig}
\usepackage[T2A]{fontenc}
\usepackage[cp1251]{inputenc}

\usepackage[eqsecnum]{cmpj2}

\issue{2015}{18}{2}{23003}
\doinumber{10.5488/CMP.18.23003}

\title[Phase separation processes in binary systems subjected to irradiation]%
{A study of phase separation processes in presence of dislocations in binary systems subjected to irradiation}%

\author[D.O. Kharchenko \textsl{et al.}]
{D.O. Kharchenko, O.M. Schokotova, A.I. Bashtova, I.O. Lysenko}
\address{Institute of Applied Physics of the National Academy of Sciences of Ukraine, \\ 58~Petropavlivska St., 40000 Sumy, Ukraine}

\date{Received November 20, 2014, in final form March 10, 2015}
\authorcopyright{D.O. Kharchenko, O.M. Schokotova, A.I. Bashtova, I.O. Lysenko, 2015}

\begin{document}

\maketitle

\begin{abstract}
Dislocation-assisted phase separation processes in binary systems subjected to
irradiation effect are studied analytically and numerically. Irradiation is
described by athermal atomic mixing in the form of ballistic flux with
spatially correlated stochastic contribution. While studying the dynamics of domain size
growth we have shown that the dislocation mechanism of phase decomposition delays
the ordering processes. It is found that spatial correlations of the ballistic flux
noise cause segregation of dislocation cores in the vicinity of interfaces
effectively decreasing the interface width. A competition between regular and
stochastic components of the ballistic flux is discussed.

\pacs 05.40.Ca, 64.75 Op, 05.70 Ln
\end{abstract}

\section{Introduction}
A study of nonequilibrium phenomena observed in materials under sustained
particle or laser irradiation attains an increasing interest in modern
theoretical physics, condensed matter physics, material science, and
metallurgy. Particle or laser irradiation causes the production of structural
disorder with generation of a large amount of point defects. These defects can
organize into defects of higher dimension and stimulate the occurrence of nonequilibrium
phenomena. In recent decades, numerous experimental data have
shown that alloys under sustained irradiation can be considered as
nonequilibrium systems manifesting phase transitions, phase separation, pattern
formation with rearrangement of point defects in bulk and on a surface (see for
example,
\cite{Shulson,Russel,OM1,OM2,OM3,JohnsonOrlov,EPJB2012,UJP2013,CMPh2013,REDS2014}).
First observations of ordering/dis\-ordering processes in irradiated alloys
were discussed six decades ago (see reference \cite{Siegel}). It was shown that
nonequilibrium conditions in such systems are caused by interactions of high
energy particles with atoms of a target (pure material, alloys).

From practical viewpoint, a study of these phenomena remains an urgent problem
to predict the behavior of construction materials. A study of phase stability in
various solids and metallic alloys under sustained irradiation received
a long-standing attention due to its intrinsic interest and its relevance in
technological problems such as: improvement of mechanical properties, radiation
resistance, radiation damage, etc. Mechanical stability of construction
materials is governed by rearrangement of the defects produced by irradiation and
their segregation on phase interfaces and grain boundaries. Perturbation of the
atomic configuration by irradiation causes the alternation of the phase stability
\cite{Martin,VaksKamyshenko,Abromeit96,MartinBellon97}. Therefore, in order to
predict the behavior of irradiated materials at different loading, one should know
the physical mechanisms leading to self-organization of the defect structure that causes microstructure transformations.

It is known that phase transformations in alloys subjected to particle
irradiation can be quite different from that observed in the absence of
irradiation \cite{Was}. Experimental observations of phase separation (spinodal
decomposition) in binary alloys (Ni-Cu, Ni-Cu) have shown that the electron
irradiation can increase the solute mobility. It causes phase decomposition at
temperatures at which diffusivities under thermal conditions are too small to
provide this effect (see, for example, references \cite{Wagner,Garner}). The same
results were obtained for alloys Au-Ni, Cu-Ni, Fe-Mo with different
compositions \cite{Nak89,Nak91_1,Nak91_2}. Irradiation damage can lead to
precipitate dissolution and stagnation precipitate ordering
\cite{Stag1,Stag2,Stag3}. It was found that phase separation in irradiated
systems can occur at temperatures above a coherent spinodal. This effect can be
described by point defects production with an increase of their mobility to
dislocations in such a way that the misfit dislocations move with the composition field
relieving strains. The corresponding model of a mobile dislocation density
field coupled with the composition field was proposed in
references \cite{Haa02,Haa04,Haa05}. It was found that phase separation is possible
above the coherent spinodal due to the motion of dislocations with a decrease of
misfit strains. Phase decomposition and patterning sustained by the dislocation
field dynamics coupled with the composition field in binary systems under sustained
irradiation were studied in reference \cite{HoytHaataja}. In this work, the irradiation
effect was considered as an additional contribution to the free energy
according to the model proposed in
references \cite{EnriqueBellon9902,EnriqueBellon00,EnriqueBellon02,EnriqueBellon04}.
In this model, the irradiation induced atomic mixing was described by a ballistic
(athermal) flux responsible for the production of a structural disorder. In
reference \cite{HoytHaataja}, the authors have shown that stable patterns characterized by
time independent amplitude and wave-length emerge due to misfit dislocations.
These linear defects are capable of reducing the coherency strain emergent at an atomic
size mismatch. In the proposed model, the  authors consider a deterministic case, where
irradiation increases the free energy of the system. A problem related to
nonequilibrium effects induced by fluctuations of the point defect concentration
and local temperature in cascades was not solved for the system with mobile
dislocations.

In this work, we extend the above model of phase separation with a dislocation
mechanism in binary systems subjected to irradiation taking into consideration stochastic
conditions. The goal of the paper is to study the role of the above fluctuation
effects in the prototype model of binary systems undergoing phase separation
assisted by mobile dislocations. We take into account the stochastic component of
the ballistic flux proposed in reference \cite{Yanovski}. Such a stochastic model was
exploited to study phase decomposition processes
\cite{EPJB2010,UJP2010,PhysA2010} and patterning \cite{CEJP2011} in irradiated
systems. The corresponding stochastic contribution takes care of local
fluctuations in the composition field due to stochasticity of point defects
concentration and temperature. We consider spatial correlations of these
fluctuations and study the effect of spatial correlations onto phase
decomposition processes. By taking into account the difference in time scales for
composition and dislocation density fields we, initially, consider the simplest
case of one slow mode (composition field). This allows us to perform the mean-field
analysis for the slow mode and study the effect of the dislocation mechanism
strength onto phase decomposition processes. The dynamics of the coupled,
simultaneous time evolved composition and dislocation fields is studied
numerically. Here, we discuss the statistical properties of phase separation and
the domain size growth law. We show that the growth of domain sizes is delayed by
dislocations participating in phase decomposition processes under sustained
irradiation. This effect was predicted theoretically in unirradiated systems
(see references \cite{DisloMech,DisloMech2}). Considering the segregation of
dislocations in the vicinity of interfaces, we discuss a competition between
regular and stochastic components of ballistic flux fluctuations.

The work is organized as follows. In section~\ref{sec:2} we present the stochastic model
of a binary system with ordinary thermal fluctuations representing the internal noise
and ballistic flux fluctuations playing the role of external noise. In section~\ref{sec:3},
we study a reduced model, where dislocation density is excluded according to
an adiabatic elimination procedure. In section~\ref{sec:4}, we numerically consider the dynamics of the phase decomposition. We conclude in section~\ref{sec:5}.

\section{Model}\label{sec:2}

Considering a binary alloy A-B, one can exploit the Bragg-Williams theory, where
the corresponding free energy density is $f_\textrm{BW}(c_\textrm{A})=Zw_\textrm{o}c_\textrm{A}c_\textrm{B}/2+T c_i\ln
c_i$. Here, $i=\{{\rm A,B}\}$, $c_i=N_i/N$, $N=N_\textrm{A}+N_\textrm{B}$ is the total number of
particles, $Z$ is a coordination number, $T$ is the temperature measured in
energetic units, an ordering energy
$w_\textrm{o}=(2w_{\{A,B\}}-w_{\{A,A\}}-w_{\{B,B\}})$ is defined through pair
interaction energies $w_{\{\cdot,\cdot\}}$. After expanding $f_\textrm{BW}$ around the
critical concentration $\overline{c}=1/2$, we arrive at Landau-like potential
$f(\psi)\simeq -A \psi^2/2+B\psi^4/4$ with
$A=T/[\overline{c}(1-\overline{c})]-Zw_\textrm{o}$ and
$B=[Zw_\textrm{o}-1/(1-\overline{c})^3+1/\overline{c}^3]/3$; the quantity $\psi$
measures the deviation from the critical concentration, i.e.,
$\psi\equiv(c-\overline{c})$. Taking into account the inhomogeneity of the alloy
and assuming that $\psi(\mathbf{r})$ varies slowly on the scale of lattice
parameter $a_0$, i.e., $\psi(\mathbf{r}+\mathbf{a}_0)\simeq
\psi(\mathbf{r})+\mathbf{a}_0\cdot\nabla \psi(\mathbf{r})$, one can take into
account the gradient energy term to the free energy in the form $r_0^2(\nabla
\psi)^2/2$, where $r_0$ is the interaction radius determining the interface width
between two phases enriched by atoms A and atoms B. Following
the Krivoglaz-Clapp-Moss expression, one has $r_0^2={\rm d}v(k)/{\rm d}k^2$, where
$v(k)$ is the Fourier transform of the atomic interaction energy. Adopting the
Cahn-Hilliard approach, the dimensionless free energy functional assumes the
Ginzburg-Landau form \cite{CahnHill,Martin90} $\mathcal{F}_0=\int{\rm
d}V\Big[-\frac{A}{2}\psi^2+\frac{B}{4}\psi^4+\frac{r_0^2}{2}(\nabla
\psi)^2\Big]$. The case $A<0$ corresponds to temperatures above the chemical
spinodal.

An additional contribution to the free energy $\mathcal{F}_0$ is given by
a lattice mismatch in the form of elastic energy $\mathcal{F}_\textrm{e}=({\nu^2
E}/{2})\int{\rm d}V\psi^2$ \cite{HoytHaataja}. Here, $E$ is the Young modulus,
$\nu$ relates to the lattice parameter change with respect to the composition
(Vegard's law), i.e., $a=a_0(1+\nu \psi)$ \cite{Cahn1961,Hoyt}. The elastic
contribution shifts the corresponding coherent spinodal: $A=\nu^2E$.

Following reference \cite{HoytHaataja}, we take into account the dislocation-assisted
mechanism for spinodal decomposition by introducing dislocation-dislocation
interactions in the form $\mathcal{F}_\textrm{d}=\!\int\!\!{\rm
d}V\!\left[\frac{C}{2}|b|^2+\frac{1}{2E}(\nabla^2\varpi)^2\right]$. Here, the constant $C$
relates to a core energy of dislocations. The elastic strain energy of the
system is governed by the Airy stress function $\varpi$ satisfying the equation
$\nabla \varpi=E(\nabla _xb_y-\nabla_yb_x)$, where $b_x$ and $b_y$ are the
corresponding components of the continuous dislocation density field in a two
dimensional problem. This term accounts for the nonlocal elastic interaction
between dislocations.

To describe the coupling between the composition field $\psi$ and strain field
of dislocations, we use the results of the work reference \cite{HoytHaataja} and introduce
a relevant contribution to the free energy in the form
$\mathcal{F}_\textrm{c}=\nu\int{\rm d}V \psi\nabla^2\varpi$. This two-dimensional
model was previously used to study the melting \cite{32,33,34}, dislocation
patterning \cite{35} and phase separation in the misfitting binary thin films
\cite{36}.

By combining all the above contributions, the total free energy of the actual
system reads
$\mathcal{F}_\textrm{tot}=\mathcal{F}_0+\mathcal{F}_\textrm{e}+\mathcal{F}_\textrm{d}
+\mathcal{F}_\textrm{c}$.
Therefore, the dynamics of the conserved fields $\psi$, $b_x$ and $b_y$ is
described by the following set of deterministic equations with diffusive
dynamics
\begin{eqnarray}
 &&\partial_t\psi=M\nabla^2\frac{\delta \mathcal{F}_\textrm{tot}}{\delta \psi}\,,\label{three_1}\\
 &&\partial_tb_x=\left(M_\textrm{g}\nabla_x^2+M_\textrm{c}\nabla^2_y\right)\frac{\delta \mathcal{F}_\textrm{tot}}{\delta b_x}\,,\label{three_2}\\
 &&\partial_tb_y=\left(M_\textrm{c}\nabla_x^2+M_\textrm{g}\nabla^2_y\right)\frac{\delta \mathcal{F}_\textrm{tot}}{\delta b_y}\,\label{three_3}.
\end{eqnarray}
Here, $M$ is the solute mobility, $M_\textrm{g}$ and $M_\textrm{c}$ denote the mobility for glide and
climb, respectively\footnote{This set of equations belongs to the models with
conserved dynamics according to the classification suggested by Galperin and Hohenberg
in reference \cite{Halpering}.}.

The effect of irradiation leads to an increase in the total free energy due
to ballistic mixing of atoms in cascades. One of the models allowing one to describe
these processes was proposed in reference \cite{EnriqueBellon9902}. It is based on
the introduction of the spatial coupling term relevant to ballistic exchanges under
irradiation conditions. The related Langevin dynamics with the additive external
noise mimicking a stochastic ballistic mixing was studied in
reference \cite{EnriqueBellon04}. It should be noted that this approach does not properly take
into account the fluctuations of the solute by a stochastic motion of the defects in
cascades. As far as these fluctuations occur in a correlated medium (crystals),
the corresponding spatial correlations of fluctuations should be
considered. The other concept of a ballistic mixing describing the above mentioned
fluctuations was proposed in
reference \cite{Martin,VaksKamyshenko,AbrMartin99,Abromeit96}. It was shown that
a ballistic mixing is stochastic in nature since the knocked atoms move at random at
the distance $R$. According to this approach, the ballistic mixing can be
described by introduction of the ballistic diffusion flux with a fluctuating
ballistic diffusion coefficient. These fluctuations are induced by irradiation
(fluctuations in both concentration of defects and local temperature in
cascades). In reference \cite{Yanovski} it was shown that such an approach leads to a
multiplicative noise Langevin dynamics, where spatially correlated external
fluctuations promote the solute flux opposite to the ordinary diffusion flux. The phase
decomposition of binary systems under the above assumptions was studied in
references \cite{EPJB2010,UJP2010}, while patterning processes in one component crystalline
systems under the irradiation effect were discussed in
references \cite{PhysA2010,CEJP2011}.

In this paper we exploit the model of a stochastic ballistic flux according to
discussions provided in
references \cite{Martin,AbrMartin99,Abromeit96,Yanovski,EPJB2010,UJP2010,CEJP2011}.
We assume that the force mixing induced by ballistic jumps occurs with
relocation distances $b\equiv\langle R\rangle=\int Rw(R){\rm d}R$, where $R$ is
distributed according to the known distribution $w(R)$. Such ballistic jumps
can be considered as a non-thermal diffusion process with a ``diffusion
coefficient'' $D^0_{b}$. The corresponding ballistic flux is
$\mathbf{J}_{b}=-D^0_{b}\nabla \psi$ \cite{Martin}. Following
reference \cite{Yanovski}, we assume that such a diffusion occurs in the fluctuating
environment. Indeed, collision processes of an energetic particle with an atom
result in local fluctuations in the temperature and a number of point defects
(Frenkel pairs). It allows one to introduce fluctuations of the ballistic flux
assuming $D^0_b\to D^0_{b}(\mathbf{r},t)+\zeta(\mathbf{r},t)$, where
$\zeta(\mathbf{r},t)$ is the random source. Therefore, the quantity
$\mathbf{J}_{b}$ has regular (deterministic) and stochastic contributions,
i.e., $\mathbf{J}_{b}=\mathbf{J}^\textrm{det}_{b}+\mathbf{J}^\textrm{stoch}_{b}$. The regular
part, $\mathbf{J}^\textrm{det}_{b}$, is characterized by the quantity
$D_{b}=\phi\sigma_r b^2$ defined through a frequency of atomic jumps
$\phi\sigma_r$, where $\phi$ and $\sigma_r$ are irradiation flux and
replacement cross-section, respectively. The corresponding stochastic part,
$\mathbf{J}^\textrm{stoch}_{b}$, relates to fluctuations in atomic relocation
distances. It is characterized by a dispersion $\langle(\delta R)^2\rangle$.
Therefore, for the ballistic flux, one can write
\begin{equation}\label{Jb}
\mathbf{J}_{b}= -\left[D_{b}+\zeta(\mathbf{r},t)\right]\nabla \psi.
\end{equation}
Here, we assume that realizations $\zeta(\mathbf{r},t)$ are independent in time
but correlated in space. Statistical properties of the external noise
$\zeta(\mathbf{r},t)$ are as follows: $\langle\zeta(\mathbf{r},t)\rangle=0$,
$\langle\zeta(\mathbf{r},t)\zeta(\mathbf{r}',t)\rangle=\sigma^2D_bC(\mathbf{r}-\mathbf{r}')\delta(t-t')$.
Here
$C(\mathbf{r}-\mathbf{r}')=(\sqrt{2\pi}r_\textrm{c})^{-2}
\exp\left\{-(\mathbf{r}-\mathbf{r}')^2/{2r_\textrm{c}^2}\right\}$
is the spatial correlation function with the correlation radius $r_\textrm{c}$;
$\sigma^2$ is an external noise intensity describing a dispersion of the
quantity $D^0_{b}$. The quantity $D_b$ in the correlator
$\langle\zeta(\mathbf{r},t)\zeta(\mathbf{r}',t)\rangle$ means that external
fluctuations are possible only at nonzero irradiation flux. In such a case, the
right-hand side of equation (\ref{three_1}) can be written as a sum of thermally
sustained diffusion flux $\mathbf{J}=-M\nabla\delta F[\psi]/\delta \psi$ and
the ballistic flux $\mathbf{J}_{b}$.

To proceed, we act onto equation (\ref{three_2}) by the operator $\nabla_y$ and act
onto equation (\ref{three_3}) by $\nabla_x$. Adding these two equations, we arrive at
one equation for the density field $\phi\equiv\nabla^4\varpi$\footnote{As far
as $\phi$ is defined in terms of gradients of $b_x$ and $b_y$, we can monitor
the strain energy reduction at segregation of dislocations at interfaces.}. In our
consideration, we take into account that the solute mobility $M$ can depend on the
field $\psi$ as $M=M_0\tilde M(\psi)$. Next, let us move to dimensionless
quantities: $\psi'=\sqrt{{B}/{A}}\psi$, $\varpi'=\sqrt{{B}/{Er_0^4}}\varpi$,
$\alpha=\sqrt{{E}/{A}}\nu$, $\mathbf{r}'=\sqrt{{A}/{r_0^{2}}}\mathbf{r}$,
$t'=({M_0A^{2}}/{r_0^{2}})t$, $D_{b}'=D_{b}/{M_0A}$,
$\mathbf{b}'=\sqrt{EBr_0^{2}/A^{3}}\mathbf{b}$,
$M_\textrm{c,g}'=M_\textrm{c,g}r_0^{2}E/M_0A^{2}$, $M_\textrm{c}=M_\textrm{g}$, $e=AC/Er_0^{2}$, $m\equiv
M_\textrm{c,g}r_0^2E/M_0A^2$. Considering a general case, let us put $\tilde
M(\psi')=1-\psi'^2$. Using the above renormalizations and dropping the primes, we arrive
at a system of two equations
\begin{equation}\label{xphi}
\begin{split}
 &\partial_t\psi=\nabla\cdot \tilde M(\psi)\left[\partial^2_{\psi\psi}\Omega(\psi)\nabla \psi-\nabla^3\psi\right]+\alpha\phi+\nabla\cdot\zeta(\mathbf{r},t)\nabla \psi+\nabla\cdot\sqrt{M(\psi)}\xi(\mathbf{r},t),\\
 &\partial_t\phi=-m\left(\phi+\alpha\nabla^2
 \psi-e\nabla^2\phi\right)
 ,
 \end{split}
\end{equation}
where
\begin{equation}
\partial_{\psi\psi}^2\Omega(\psi)=\partial_{\psi\psi}^2f(\psi)+\frac{D_b}{\tilde M(\psi)}\,,\qquad
f(\psi)=-\frac{1-\alpha^2}{2}\psi^2+\frac{\psi^4}{4}\,.
\end{equation}
The last term in the equation for $\psi$ represents an internal multiplicative
noise. It is characterized by $\langle\xi\rangle=0$ and $\langle
\xi(\mathbf{r},t)\xi(\mathbf{r},t)\rangle=\theta
\delta(\mathbf{r}-\mathbf{r}';t-t')$, where $\theta$ is the parameter measuring
the internal noise intensity proportional to a bath temperature. We assume that
no spatio-temporal correlations between fluctuation sources are possible.

It should be noted that time scales of the evolution of composition and dislocation
density fields described by the quantity $m\propto M_\textrm{c,g}/M_0$ can be
different. At $m=0$, we get a system with immobile dislocations. Limit $m\to
\infty$ corresponds to extremely mobile dislocations. It means that
$m\in[0,\infty)$ depends on the properties of the studied material and can be
considered as a free parameter of the model. A detailed study of the systems
characterized by different values of $m$ was reported in
reference \cite{AppPhysLett2005}. Next, following reference \cite{AppPhysLett2005}, we
consider the system with mobile dislocations by taking $m\geqslant 1$. In the
simplest case of extremely mobile dislocations ($m\gg 1$), one can
adiabatically eliminate the fast field considering the behavior of the slow one. In our
further study, we discuss statistical properties of the system according to
subordination principle. To make a general analysis, we study the behavior of
the system with the above two fields by taking into account the above time scales
difference.

\section{Subordination principle and mean-field results}\label{sec:3}
\subsection{Stability of the reduced system}
Let us consider the simplest case when mobile dislocations instantaneously
adjust the evolving composition field. To this end, we put $m\gg 1$. This allows
us to exclude the fast variable $\phi$ by assuming $m^{-1}\partial_t\phi\simeq
0$. Hence, using the Fourier representation for the Fourier components $\phi_k$
and $\psi_k$, we obtain the relation $\phi_k\simeq \frac{\alpha
k^2}{1+ek^2}\psi_k$ from the second equation of the system (\ref{xphi}). In the
case $ek^2\ll 1$, we can expand the denominator up to the first order and obtain
an approximation $\phi_k\simeq \alpha k^2(1+ek^2)\psi_k$, or $\phi\simeq
-\alpha \nabla^2(1-e\nabla^2)\psi$. Substituting this expression into the first
equation of the system (\ref{xphi}), we get one equation for the slow mode in
the form
\begin{equation}\label{xeq}
 \partial_t\psi=\nabla\cdot \tilde M(\psi)\nabla\tilde\mu(\psi)+
 \nabla\cdot\left[\zeta(\mathbf{r},t)\nabla
 \psi+\sqrt{M(\psi)}\xi(\mathbf{r},t)\right],
 \end{equation}
where the notation $\nabla\tilde\mu(\psi)\equiv
\nabla\mu_\textrm{ef}(\psi)-\frac{\alpha^2}{\tilde M(\psi)} \nabla(1-e\nabla^2)\psi$
is introduced for convenience; $\mu_\textrm{ef}(\psi)$ plays the role of the effective
chemical potential [$\nabla \mu_\textrm{ef}=\partial^2_{\psi\psi}\Omega(\psi)\nabla
\psi-\nabla^3\psi$]. The obtained equation (\ref{xeq}) is the main equation for
the reduced system analysis. According to the structure of equation (\ref{xeq}), one
should have in mind that for the field $\psi$ we get conserved dynamics, i.e.,
$\int \psi(\mathbf{r},t){\rm d}\mathbf{r}=\psi_0$, where $\psi_0$ stands for the
initial concentration difference; $\psi_0=\textrm{const}$ according to the mass conservation
law.

In statistical analysis we study only observable (averaged) quantities. By
averaging equation (\ref{xeq}) one gets noise correlators which can be calculated
using the Novikov's theorem \cite{Novikov} (the corresponding averatging procedures
are shown in appendix~A). The thermal flux (internal) noise correlator reads: \linebreak
$\nabla\cdot \langle\sqrt{\tilde M}\xi\rangle=-(\theta/2)\nabla\cdot
\langle\nabla
\partial_\psi \tilde M\rangle$.
Calculations for the external noise correlator give: $\langle\zeta\nabla
\psi\rangle=$ \linebreak $\sigma^2\left[C(\mathbf{0})\nabla^3\langle
 \psi\rangle+C''(\mathbf{0}) \nabla \langle
\psi\rangle\right]$, where we have to note that $C(\mathbf{r-r'})$ acquires its
maximal value at $\mathbf{r=r'}$, which implies that $\left.\nabla
C(\mathbf{r-r'})\right|_{\mathbf{r=r'}}=0$; $
\nabla^2\left.C(\mathbf{r-r'})\right|_{\mathbf{r=r'}}\equiv C''(\mathbf{0})<0$
(see references \cite{Garcia,Yanovski,UJP2008,EPJB2010} for details). Therefore,
after averaging we get
\begin{equation}\label{eq21}
\partial_t \langle \psi\rangle=\nabla\cdot \langle \tilde M\nabla\tilde\mu\rangle
 -\frac{\theta}{2}\nabla\cdot \langle\nabla \partial_\psi \tilde M\rangle+
\sigma^2\left[C''(\mathbf{0})\nabla^2\langle \psi\rangle+
C(\mathbf{0})\nabla^4\langle \psi\rangle\right].
\end{equation}

Let us study the stability of the homogeneous state $\psi=0$. As far as we consider
the system with conserved dynamics, the corresponding stability analysis can be
done studying the dynamics of the structure function $S(k,t)$ as the Fourier transform
of the two point correlation function
$\langle\psi(\mathbf{r},t)\psi(\mathbf{r}',t)\rangle$. Linearizing the system
in the vicinity of the state $\psi=0$, in the continuous and thermodynamic
limit, we arrive at the dynamical equation for the structure function in the
form (see appendix~B for details)
\begin{equation}\label{F2}
\frac{{\rm d}S(k,t)}{{\rm d}t}=-2k^2\omega(k)S(k,t)+2\theta k^2-\frac{2
k^2}{(2\pi)^2}\int{\rm d}\mathbf{q}S(q,t)
+\frac{2k^2D_{b}\sigma^2}{(2\pi)^2}\int{\rm
d}\mathbf{q}C(|\mathbf{k}-\mathbf{q}|)S(\mathbf{q},t),
\end{equation}
with the dispersion relation
\begin{equation}\label{w(k)1}
\omega(k)=\epsilon_\textrm{ef}+ \beta_\textrm{ef}k^2.
\end{equation}
Here, $\epsilon_\textrm{ef}$ is the effective control parameter playing the role of an
effective temperature counted from the critical value and $\beta_\textrm{ef}$ is the
inhomogeneity parameter defined as
\begin{equation}\label{eff}
\epsilon_\textrm{ef}=\alpha^2-1+\theta+D_{b}\left[1+ \sigma^2C''(0)\right],\qquad
\beta_\textrm{ef}=1-\alpha^2e -D_{b}\sigma^2 C(0).
\end{equation}
It follows that the ballistic diffusion (its regular component) increases the
effective temperature of the system, whereas correlation effects governed by
the term $C''(0)<0$ decrease its value. At the same time, ballistic diffusion is
capable of decreasing the interface width between two phases [last term in
 $\beta_\textrm{ef}$ in equation (\ref{eff})].

From equation (\ref{w(k)1}),  one finds that the critical wave-number that bounds the unstable
modes is defined as
\begin{equation}
k_\textrm{c}=\sqrt{\frac{1-\alpha^2-\theta-D_{b}\left[1+ \sigma^2C''(0)\right]}{1-\alpha^2e
-D_{b}\sigma^2 C(0)}}.
\end{equation}
The most unstable mode is described by the wave-number $k_\textrm{m}=k_\textrm{c}/\sqrt{2}$. For
the actual set of the system parameters at $\alpha<1$, one gets a decreasing
dependence $k_\textrm{c}(\alpha)$. Therefore, at small $\alpha$, the dislocation
mechanism promotes a decrease in the wave-number of unstable modes. With an
increase in $\alpha$, spatial modulations of the composition field are characterized
by long-wave modes. At the same time, spatial correlations of the external noise
$\zeta$ increase the wave-number of unstable modes due to $C''(0)<0$.

\begin{figure}[!t]
\centerline{
\includegraphics[width=0.43\textwidth]{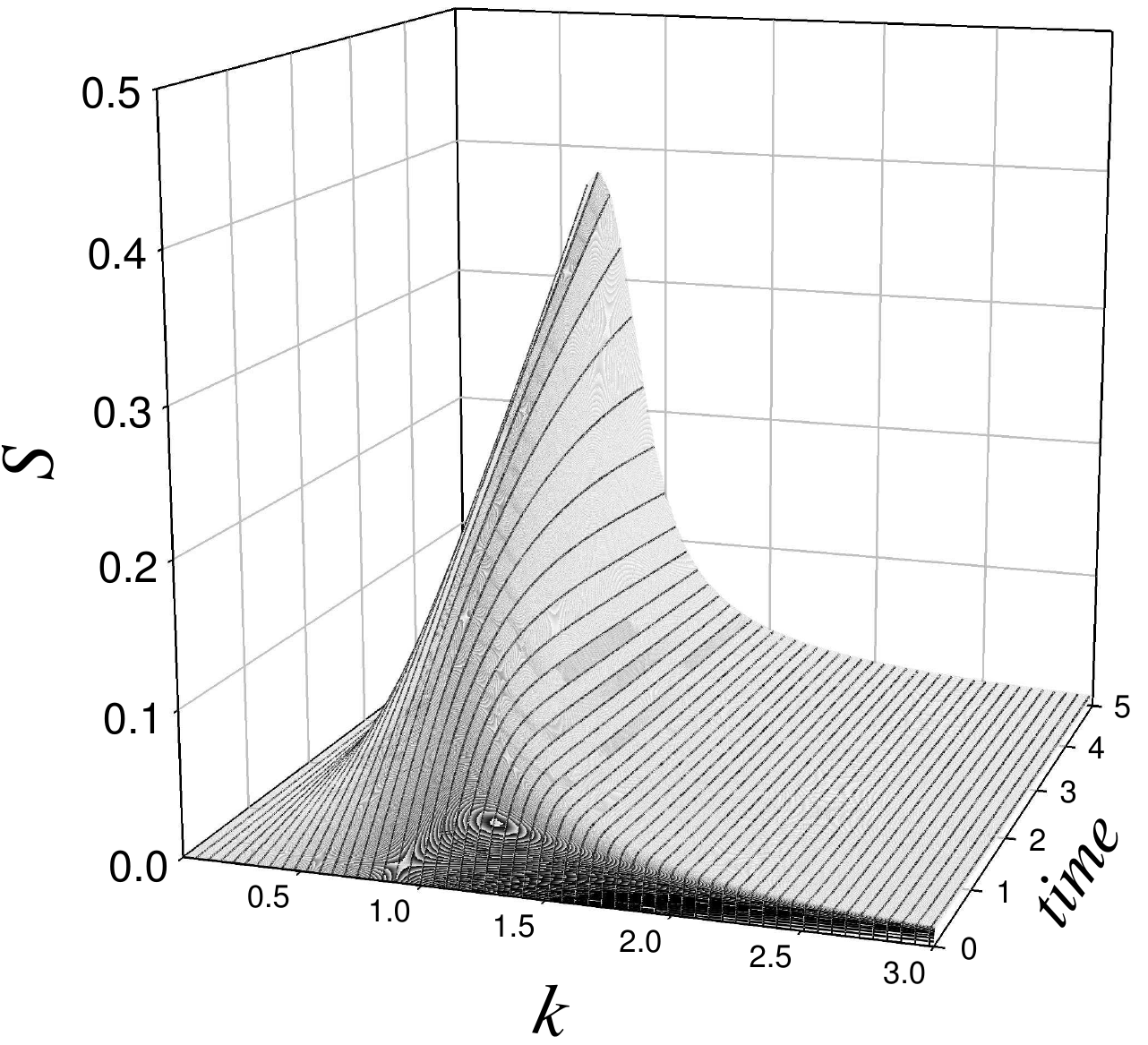}
\hspace{1cm}%
\includegraphics[width=0.45\textwidth]{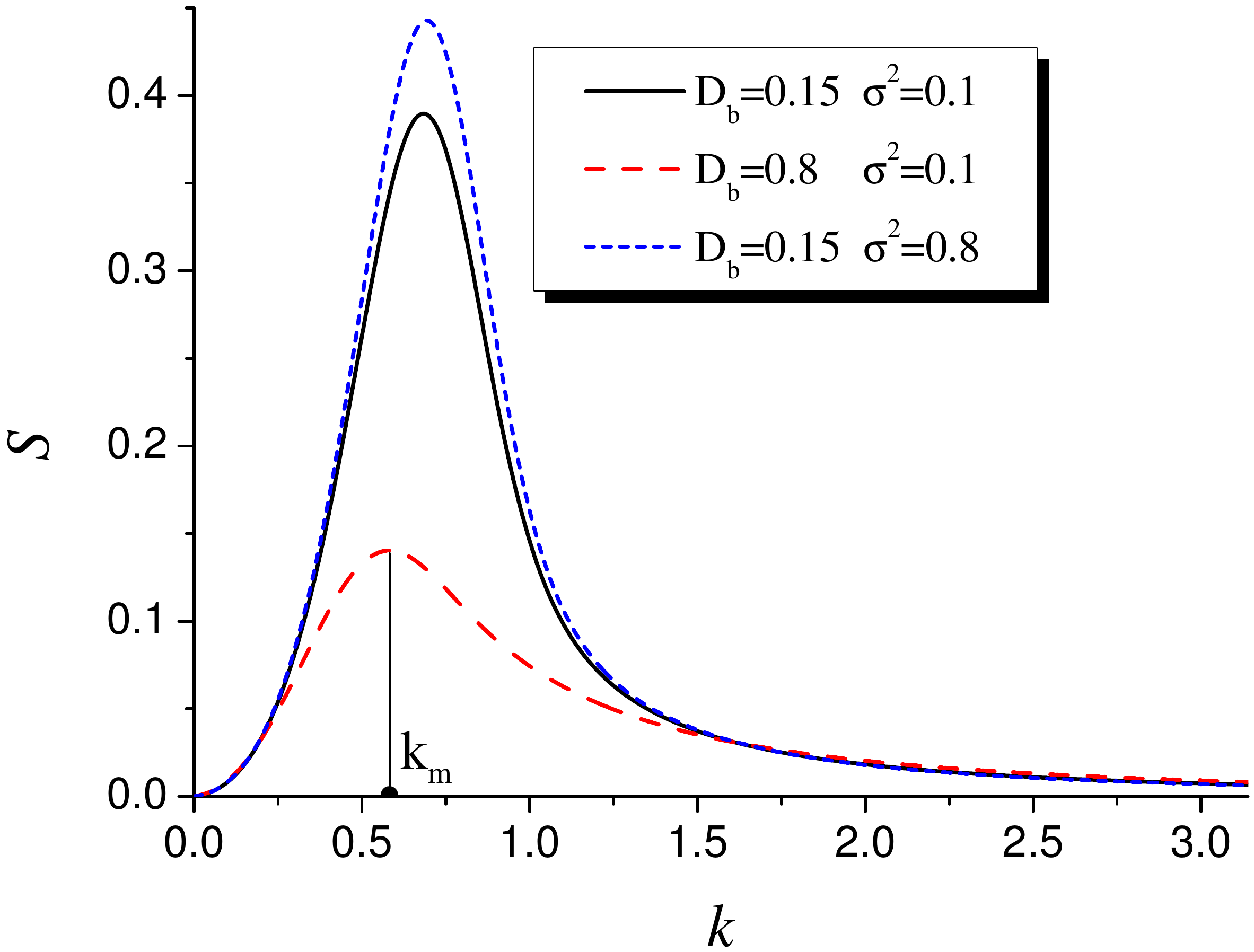}
}
\hspace{3.5cm} (a) \hspace{7.5cm} (b)
\caption{(Color online) The structure function for
the reduced system: (a) the dynamics of $S(k,t)$ at $D_b=0.15$, $\sigma^2=0.1$; (b)
the dependence $S(k)$ at a fixed time interval ($t=10$). Other parameters are:
$\alpha=0.5$, $\theta=0.1$, $e=0.2$, $r_\textrm{c}=1$.\label{figS_one}}
\end{figure}

Typical dynamics of the structure function $S(k,t)$ are shown in
figure~\ref{figS_one}~(a), $k_\textrm{m}$ relates to the position of the peak in the dependence
$S(k)$ [see figure~\ref{figS_one}~(b)]. From figure~\ref{figS_one}~(a) one can see that
during the system evolution, the peak of $S(k,t)$ related to the wave-number
$k_\textrm{m}$ moves toward $k=0$ and its height increases. Therefore, the corresponding
spatial instability promotes the ordering processes with the formation of domains of
phases enriched by atoms A or B. The effect of the system parameters onto
$S(k)$ is shown in figure~\ref{figS_one}~(b). Here, one can find that an increase in
the ballistic mixing coefficient $D_b$ promotes the  formation of the structural
disorder characterized by realization of long-wave perturbations and small
maximal value of the structure function. The stochastic contribution of the
ballistic mixing flux acts in an opposite manner stimulating the ordering processes.
At elevated values of external noise intensity $\sigma^2$, the corresponding
spatial structures are characterized by lower domain sizes enriched by the atoms of
one sort. This effect is caused by spatial correlations of the external
fluctuations.

The linear stability analysis is valid only on a short time scale. At large
time limit ($t\to \infty$), one can use the mean-field approximation based on
the analysis of the solution of the Fokker-Planck equation for the probability
density of the composition field.

\subsection{Mean-field approximation}

To analytically study the statistical properties of phase separation at $t\to \infty$
 one needs to analyze a stationary probability density
$\mathcal{P}_\textrm{s}([\psi])$. The behavior of the system can be described analytically
within the framework of the mean-field approach derived for the systems with
conserved dynamics \cite{IGTS99,GO93,GO2001,PhysA2008,EPJB2010,UJP2010}.

In the Wiess mean-field approximation, one can use the mean field (molecular field)
$\eta\equiv \langle \psi\rangle$ as an order parameter for phase transitions
and phase decomposition. In such a case, one uses the transformation procedure,
which allows us to introduce the order parameter in $d$-dimensional space as follows:
\begin{equation}
\Delta\psi\equiv\rightarrow 2d (\langle \psi\rangle-\psi).
\end{equation}
The mean-field value $\eta$ is self-consistently defined according to the
definition of the mean $\langle\psi\rangle$ through the stationary distribution
function $P_\textrm{s}$. In the mean-field theory, the stationary distribution is a function
of $\psi$ and $\eta$. A procedure to obtain the corresponding distribution as a
solution of the corresponding Fokker-Planck equation is shown in appendix~C.

In order to define the transition and critical points at phase separation, we use the
procedure proposed in references \cite{IGTS99,GO93,GMS83}. According to this approach
in a deterministic case with $D_{b}=0$, one has a model
$\partial_t\psi=\nabla\cdot M\nabla \delta \mathcal{F}/\delta \psi$, where the
restriction $\psi_0=\int_V{\rm d}\mathbf{r}\psi(\mathbf{r},t)$ is taken into
account, $\psi_0$ is fixed by the initial conditions. For such a system, the
transition point is $\Theta_\textrm{T}(\psi_0)$: at $\Theta>\Theta_\textrm{T}(\psi_0)$, the
homogeneous state $\psi_0$ is stable; at $\Theta<\Theta_\textrm{T}(\psi_0)$, the system
separates into two bulk phases, $\psi_1$ and $\psi_2$, fulfilling $\langle
\psi\rangle=\psi_0$. The transition from a homogeneous state to two-phase state
is critical for $\psi_0=0$ only, i.e. $\Theta_\textrm{T}(0)=\Theta_\textrm{c}$ is the critical
point. The corresponding steady state solutions are given as solutions of the
equation $\nabla M\nabla \delta \mathcal{F}/\delta \psi=0$. If no flux
condition is applied, then the bounded solution is $\delta \mathcal{F}/\delta
\psi=h$, where $h$ is a constant effective field (in equilibrium systems $h$ is
a chemical potential). In the homogeneous case, the value $h$ depends on the
initial conditions $\psi_0$. Above the transition point, the steady state is not
globally homogeneous. Here, the system separates into two bulk phases with
the values $\psi_1$ and $\psi_2$. In the case of the symmetric form of the free
energy functional where two phases with $\psi_1=-\psi_2$ are realized, we get
$h=0$ \cite{IGTS99}. Hence, if the field $h$ becomes trivial, then the
transition point can be defined.

By using this procedure, one finds that in the homogeneous case the mean-field
is the same everywhere and equals the initial value, i.e. $\eta=\psi_0$. Hence,
solving the self-consistency equation
\begin{equation}\label{slf}
\eta=\int\limits_{-1}^1 \psi P_\textrm{s}(\psi,\eta,h){\rm d}\psi
\end{equation}
at the fixed mean-field value, we obtain the constant effective field $h$. Below
the threshold $\Theta_\textrm{T}$, the system decomposes into two equivalent phases with
$\langle \psi_1\rangle=-\langle \psi_2\rangle$, and $h$ should be the same for
these two phases and should be zero. Hence, below the threshold only $\langle
\psi\rangle$ should be defined as a solution of the self-consistency equation
with $P_\textrm{s}(\psi,\eta,0)$.

\begin{figure}[!b]
\centerline{
\includegraphics[width=0.48\textwidth]{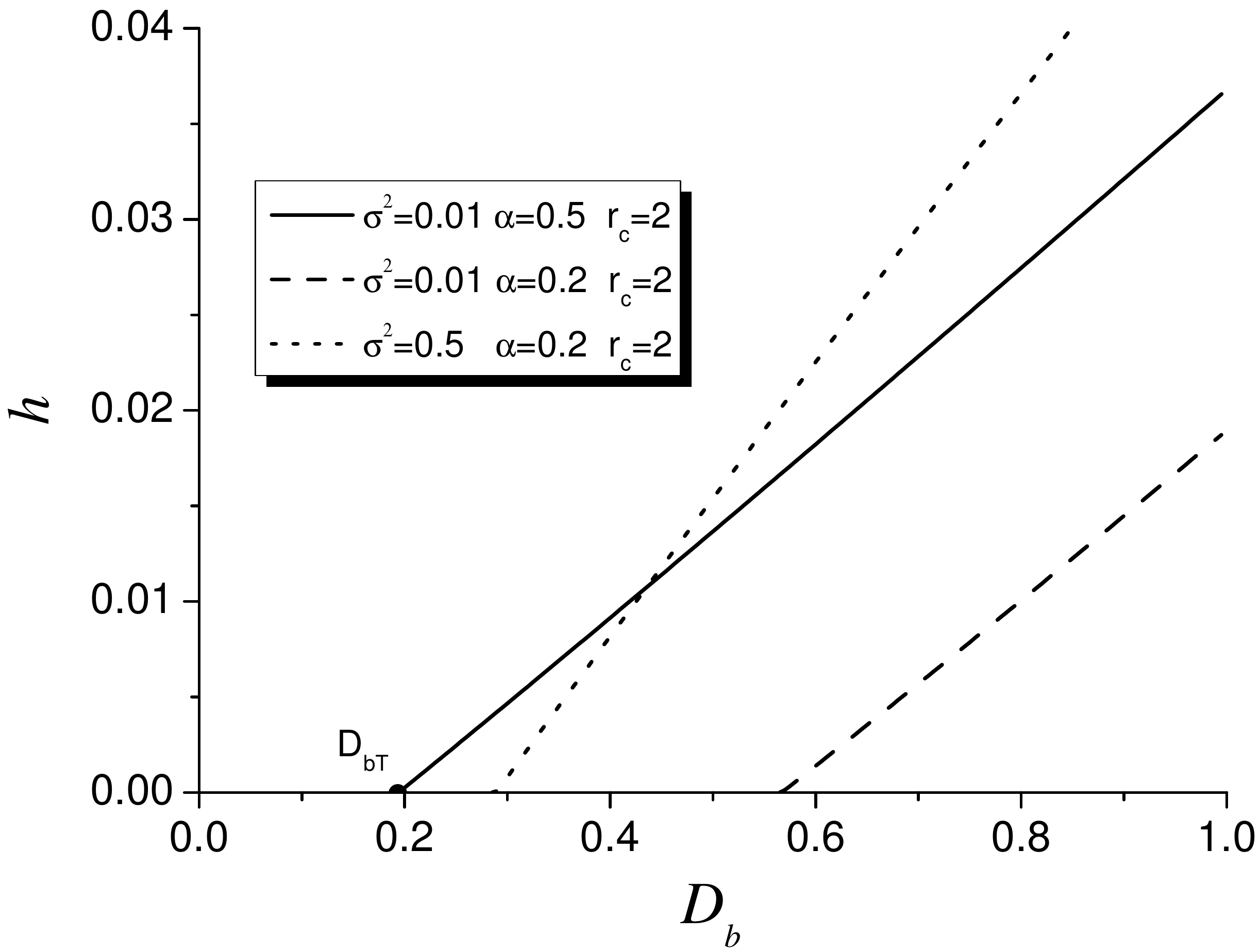}
\hspace{5mm}
\includegraphics[width=0.48\textwidth]{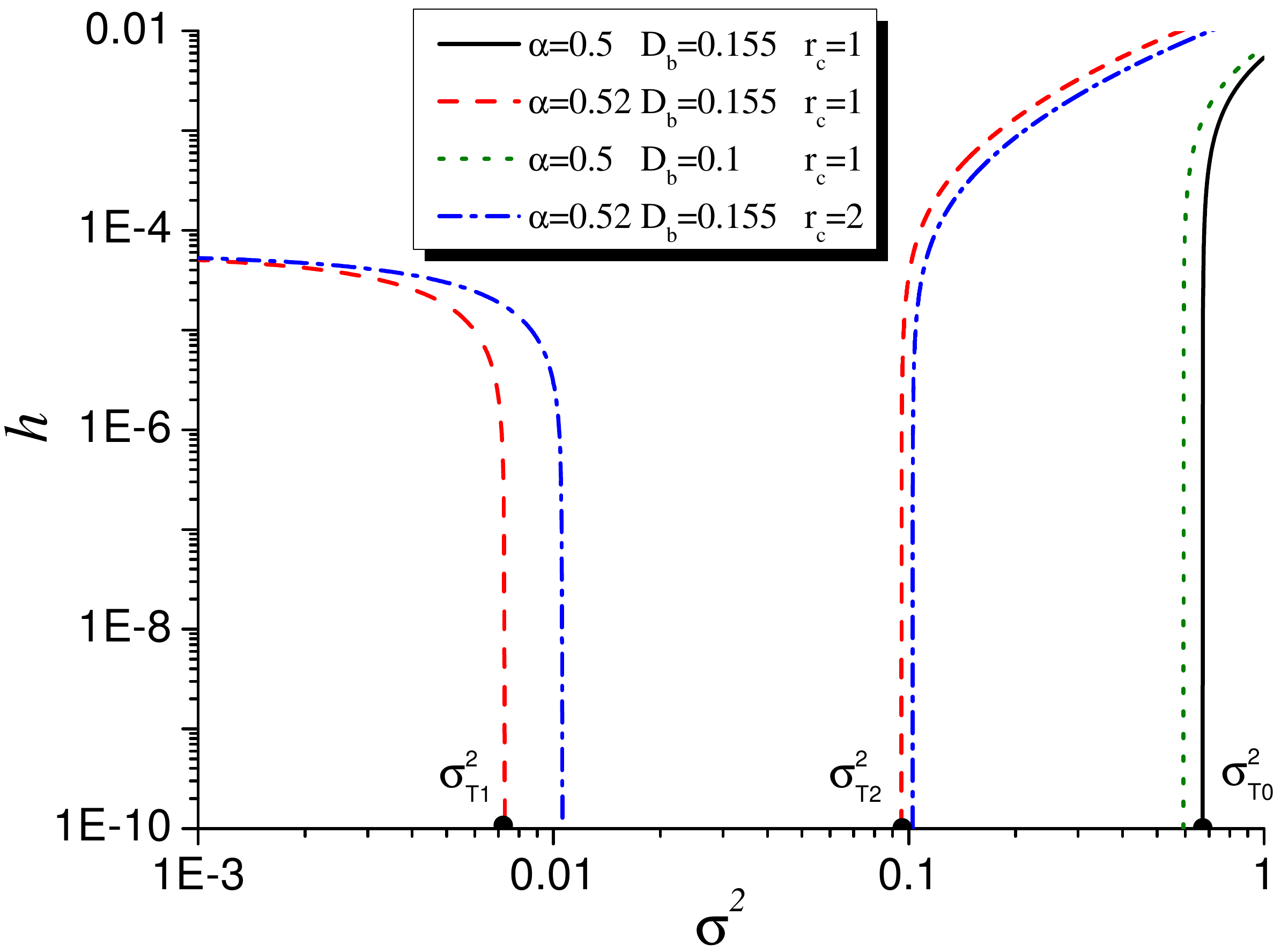}
}
\hspace{3.8cm} (a) \hspace{7.2cm} (b)
\caption{(Color online) Constant field $h$ {versus} $D_b$ and $\sigma^2$ [panels (a) and
(b), respectively] at $\psi_0=0.05$ and different sets of the system
parameters. \label{hDsigma}}
\end{figure}

In the actual case, we are interested in phase decomposition phenomena induced
by the irradiation effect. Therefore, we define the transition and critical
points only for the parameters relevant to irradiation, namely $D_b$, $\sigma^2$ by
fixing $\psi_0$. The corresponding dependencies of the effective field $h$
{versus} $D_b$ and the external noise intensity are shown in
figure~\ref{hDsigma}.
From figure~\ref{hDsigma}~(a) it is seen that the field $h$ takes nonzero values
above the transition point $D_{b\textrm{T}}$. According to the definition of $h$ as a
chemical potential, one can say that at fixed $\psi_0$, the quantity $h$ is
compensated and phase separation occurs inside the domain of the parameters
where $h=0$. From the dependencies $h(D_{b})$ it follows that the ordered state
with the initial concentration $\psi_0$ can be found only before $D_{b\textrm{T}}$. It
is seen that with an increase in the strength of the dislocation mechanism
described by $\alpha$, the transition point $D_{b\textrm{T}}$ decreases. Considering the
dependence $h(\sigma^2)$ [see figure~\ref{hDsigma}~(b)], it follows that with an
increase in $\alpha$, the phase separation processes can be realized inside the
noise intensity interval $(\sigma^{2}_\textrm{T1},\sigma^{2}_\textrm{T2})$.

\begin{figure}[!t]
\centerline{
\includegraphics[width=0.48\textwidth]{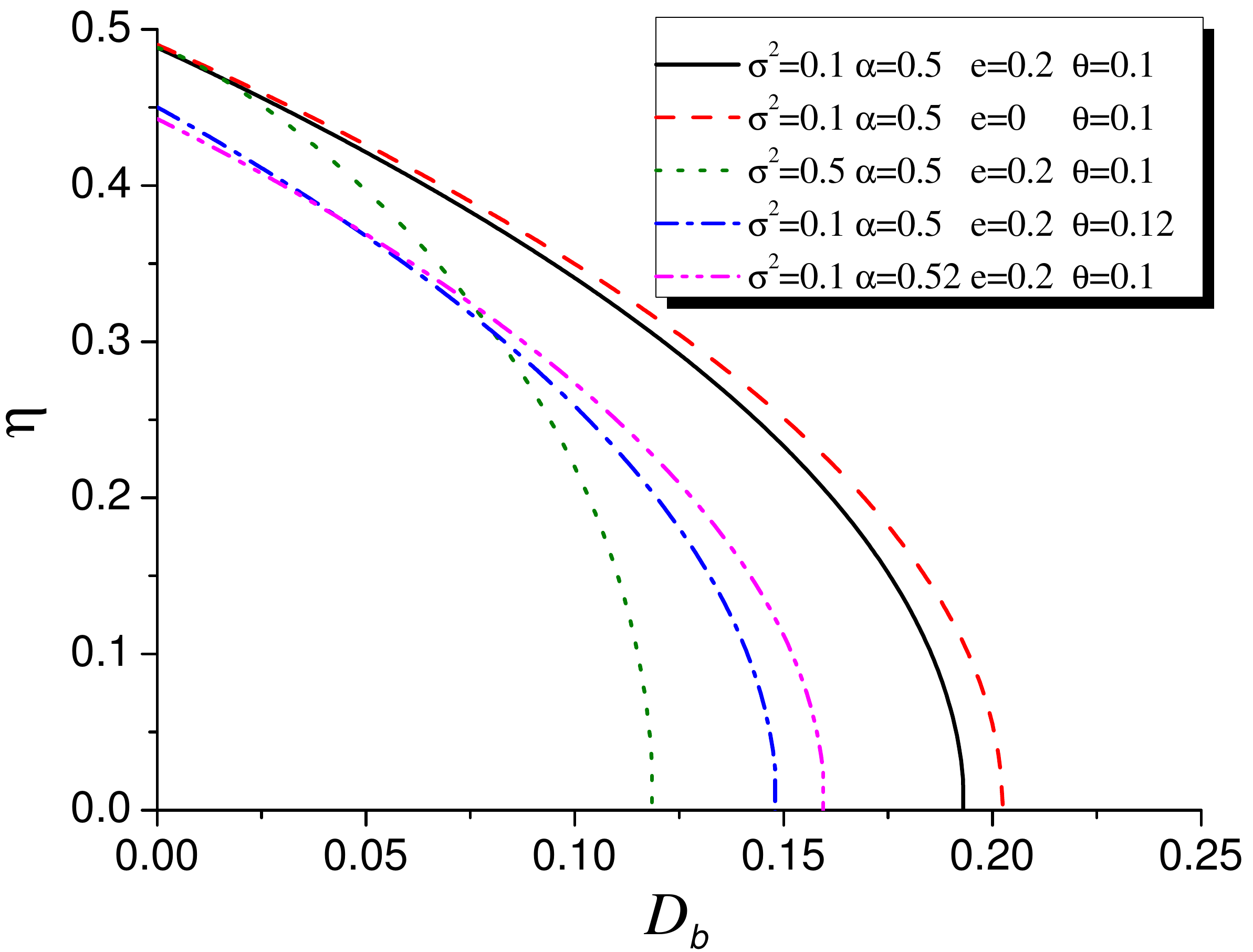}
\hspace{5mm}
\includegraphics[width=0.48\textwidth]{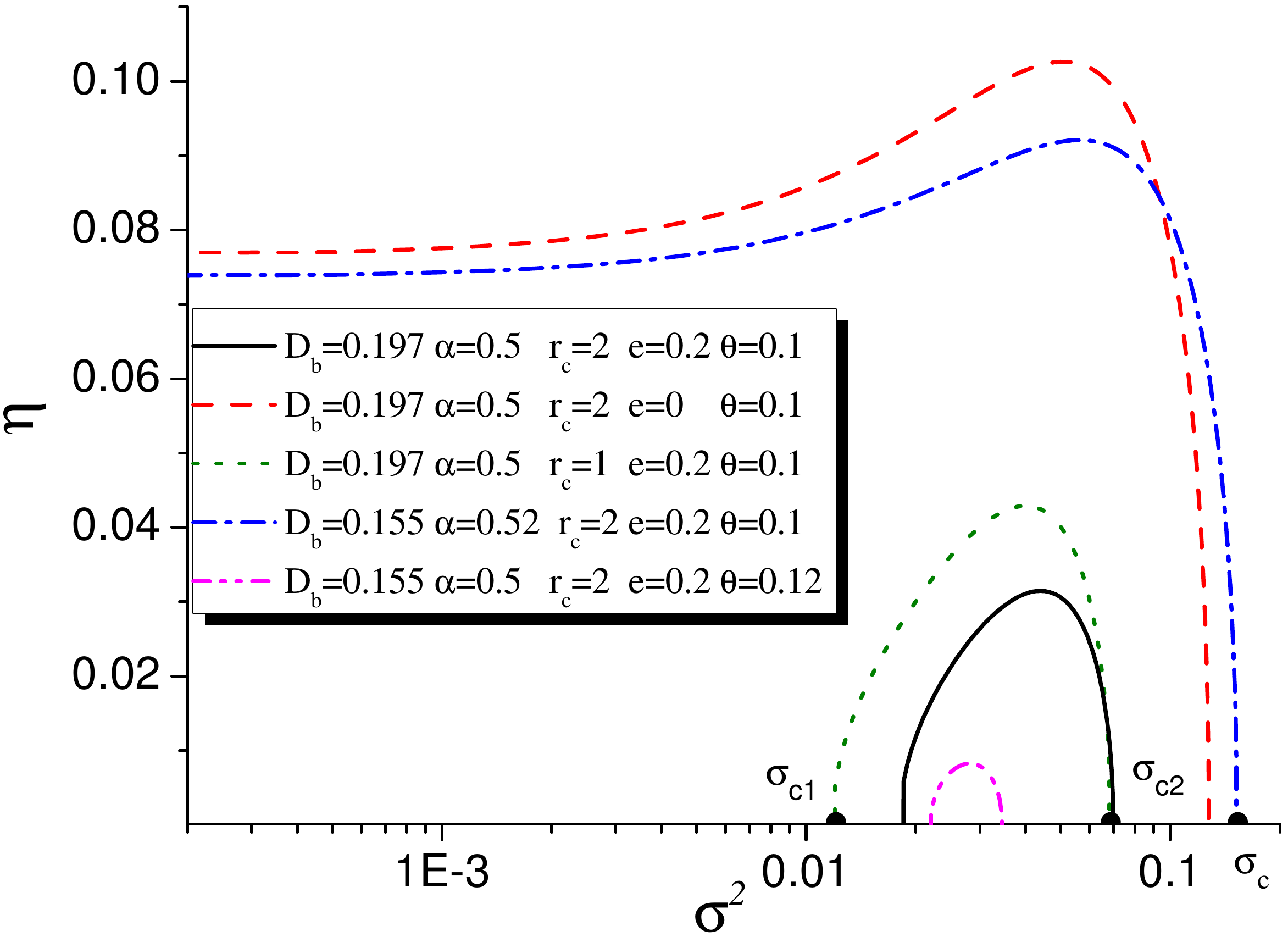}
}
\hspace{3.7cm} (a) \hspace{7.2cm} (b)
\caption{(Color online) The mean-field value $\eta$ {versus} ballistic diffusion
coefficient $D_b$ and noise intensity $\sigma^2$ [panels (a) and panel (b),
respectively].\label{eta(Dsigma)}}
\end{figure}

Next, let us discuss the mean-field $\eta$ behavior varying the system
parameters. Here, we solve the self-consistency equation at $h=0$. According to
figure~\ref{eta(Dsigma)}~(a), the mean-field decreases with an increase in the
coefficient $D_{b}$. It behaves critically in the vicinity of the value
$D_{b}=D_{b\textrm{c}}$, when nontrivial values $\langle \psi_1\rangle=-\langle
\psi_2\rangle$ appear. Here, we arrive at the conclusion that the irradiation
leads to homogenization of the composition field distribution. By increasing
the noise intensity, one finds that phase decomposition is realized at lower
values of $D_b$ compared to the case of small $\sigma^2$. An increase in the
intensity of internal fluctuations $\theta$ suppresses the phase separation at
large $D_b$. The same effect can be found when the intensity of the feedback
between dislocation density and composition field increases. This result
follows even from the analysis of the deterministic system, where the elastic field
changes the critical point position. A more interesting situation is observed by
varying the external noise intensity [see figure~\ref{eta(Dsigma)}~(b)]. Here, one
finds that the external noise leads to the emergence of a disordered state ($\eta=0$) at
$\sigma^2>\sigma^2_\textrm{c}$. In other words, external fluctuations of large
intensity lead to a statistical disorder. On the other hand, at special choice of
the system parameters related to ballistic flux properties, a reentrant behavior
of the mean-field $\eta$ is observed. Here, phase decomposition is realized in a
window of the noise intensity $\sigma^2\in[\sigma^2_\textrm{c1}, \sigma^2_\textrm{c2}]$. This
phenomenon is caused by the competition between regular and stochastic (correlated)
parts of the ballistic flux. At $\sigma^2<\sigma^2_\textrm{c1}$, the most essential
contribution is given by the regular component $D_b$ leading to homogenization
of the alloy. Inside the interval $\sigma^2\in[\sigma^2_\textrm{c1}, \sigma^2_\textrm{c2}]$,
the correlation effects dominate and lead to a decrease in the effective
temperature of the system. At large $\sigma^2$, external fluctuations destroy
the ordered state. Therefore, the correlated ballistic flux is capable of inducing phase
separation processes of initially homogeneous alloys. According to dependencies
$h(\sigma^2)$, one can conclude that the dislocation mechanism sustains the above
reentrance.

\begin{figure}[!t]
\centering
\includegraphics[width=0.5\textwidth]{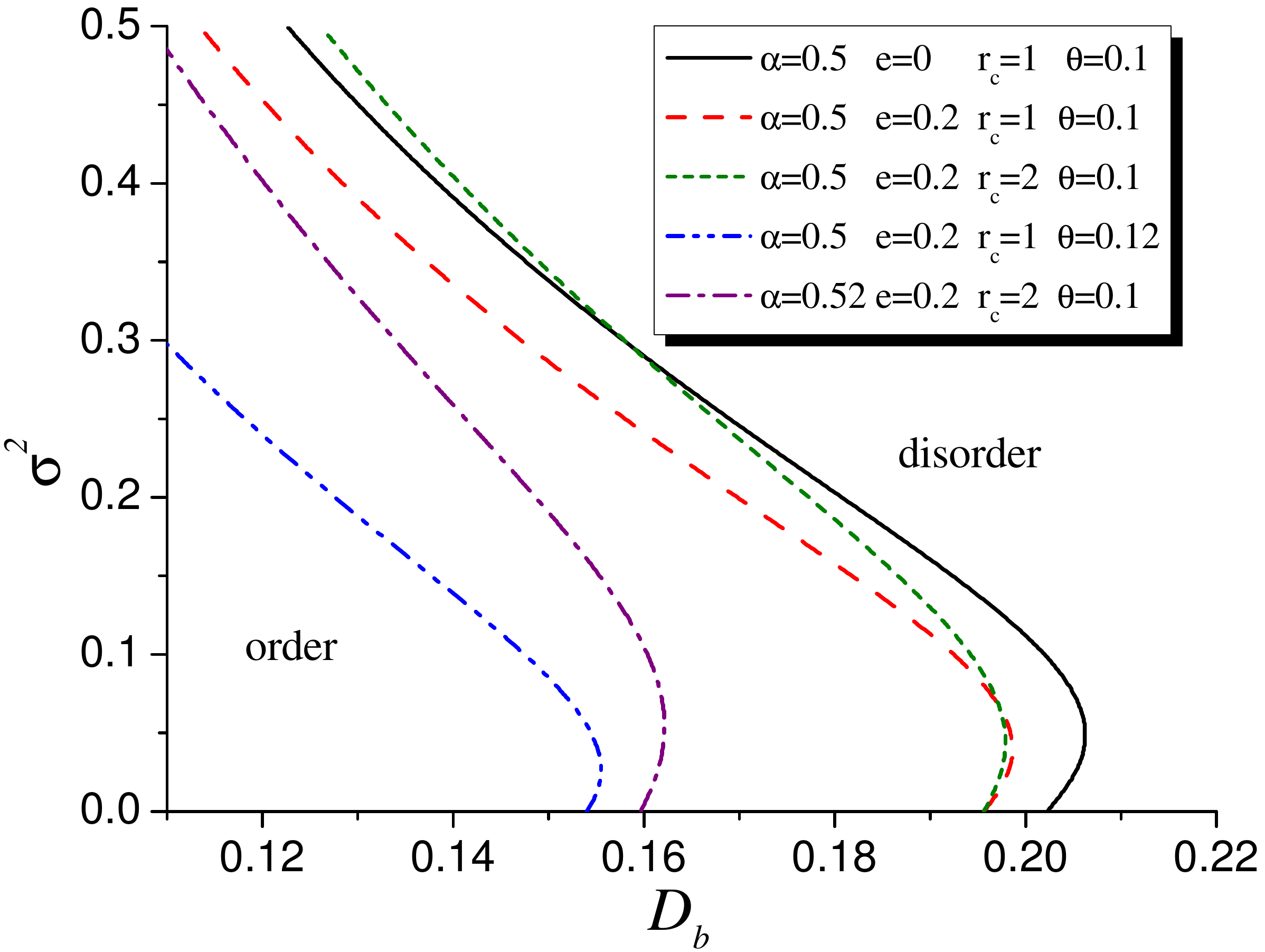}
\caption{(Color online) The mean-field phase diagram at different sets of the system
parameters.\label{phdgr}}
\end{figure}

The corresponding phase diagram illustrating the formation of ordered and
disordered phases is shown in figure~\ref{phdgr}. It is seen that a reentrant phase
separation is realized in a narrow interval for $\sigma^2$ at elevated $D_b$.
At small $D_b$, one gets the standard scenario of phase decomposition, where
fluctuations suppress the ordering processes. Comparing different curves, it
follows that an increase in the intensity of internal fluctuations $\theta$
shrinks the interval for the system parameters bounding the domain of the ordered phase
(cf. dash and dash-dot-dot curves). The domain of reentrant ordering extends
with an increase in the external noise correlation radius $r_\textrm{c}$ (cf. dash and
dot lines). At an elevated $r_\textrm{c}$, the size of the domain for the ordered phase
grows at small $D_b$. An increase in $\alpha$ shrinks the domain of the ordered
phase and extends the domain of the reentrant decomposition (cf. dot and dash-dot
curves).

According to the obtained results, it follows that phase separation processes can be
controlled by the main system parameters (temperature and elastic properties of
the alloy) and statistical properties of irradiation effect (regular and
stochastic contributions in the ballistic flux). Moreover, correlated
stochastic contribution of this flux is capable of inducing reentrant phase
separation processes.

Let us study a strong coupling limit, neglecting the spatial interactions term.
Assuming $\psi=\eta$, one gets the stationary distribution in the form
$P_\textrm{s}(\psi,\eta)=\delta(\psi-\eta)$. To obtain an equation for the effective
field $h$, we integrate equation (\ref{st_FPE})
 and find
\begin{equation}\label{scl}
h=M(\eta)\left[\partial_\eta
f(\eta)+\frac{D_{b}}{M(\eta)}\eta\right]-\frac{\theta}{2}
\partial_\eta M(\eta)-
2dD_{b}\sigma^2 (C_0-C_1)\eta.
\end{equation}
At $h=0$, one has solutions for two bulk phases
\begin{equation}\label{eta_pm}
\eta_{\pm}=\pm\frac{1}{2}\sqrt{4-2\alpha^2-2\sqrt{\alpha^4+4(D_b+\theta)-8dD_b\sigma^2(C_0-C_1)}}.
\end{equation}
The corresponding transition line can be obtained directly from
equation (\ref{eta_pm}) at $\eta=\psi_0$, where $\psi_0$ is the initial value for the
composition field. Critical values for the system parameters can be obtained
from the condition $\eta=0$. It is interesting to note that ballistic flux
parameters lead to renormalization of the effective temperature: $\Theta$
counting off the critical one $\Theta_\textrm{c}=1$. Indeed, according to the definition
of the free energy density $f(\psi)$ for an unirradiated system, the quantity
$\Theta$ is reduced to $\alpha^2$. In an irradiated system, $\Theta$ is reduced to
$\alpha^2+D_b+\theta-2dD_b\sigma^2(C_0-C_1)$. Hence, the regular component of
the ballistic flux increases the effective temperature in the same manner as
the internal noise does. On the other hand, the external fluctuations reduce
this temperature due to their spatial correlations. From equation (\ref{eta_pm}) it
follows that an increase in $D_b$, $\theta$, and $\alpha$ causes a decrease
in the order parameter. The external noise is capable of extending the interval for
$D_b$ where the mean-field takes up nonzero values.

It is known that the mean-field results are mostly qualitative. To validate
the mean-field results, we will use a simulation procedure. In the next section
we discuss the behavior of our system considering the dynamics of both quantities
$\psi$ and $\phi$ and numerically illustrate a possibility of reentrant phase
separation processes.

\section{The effect of dislocation density field dynamics}
\label{sec:4}

\subsection{Stability analysis}
\begin{figure}[!t]
\centerline{
\includegraphics[width=0.48\textwidth]{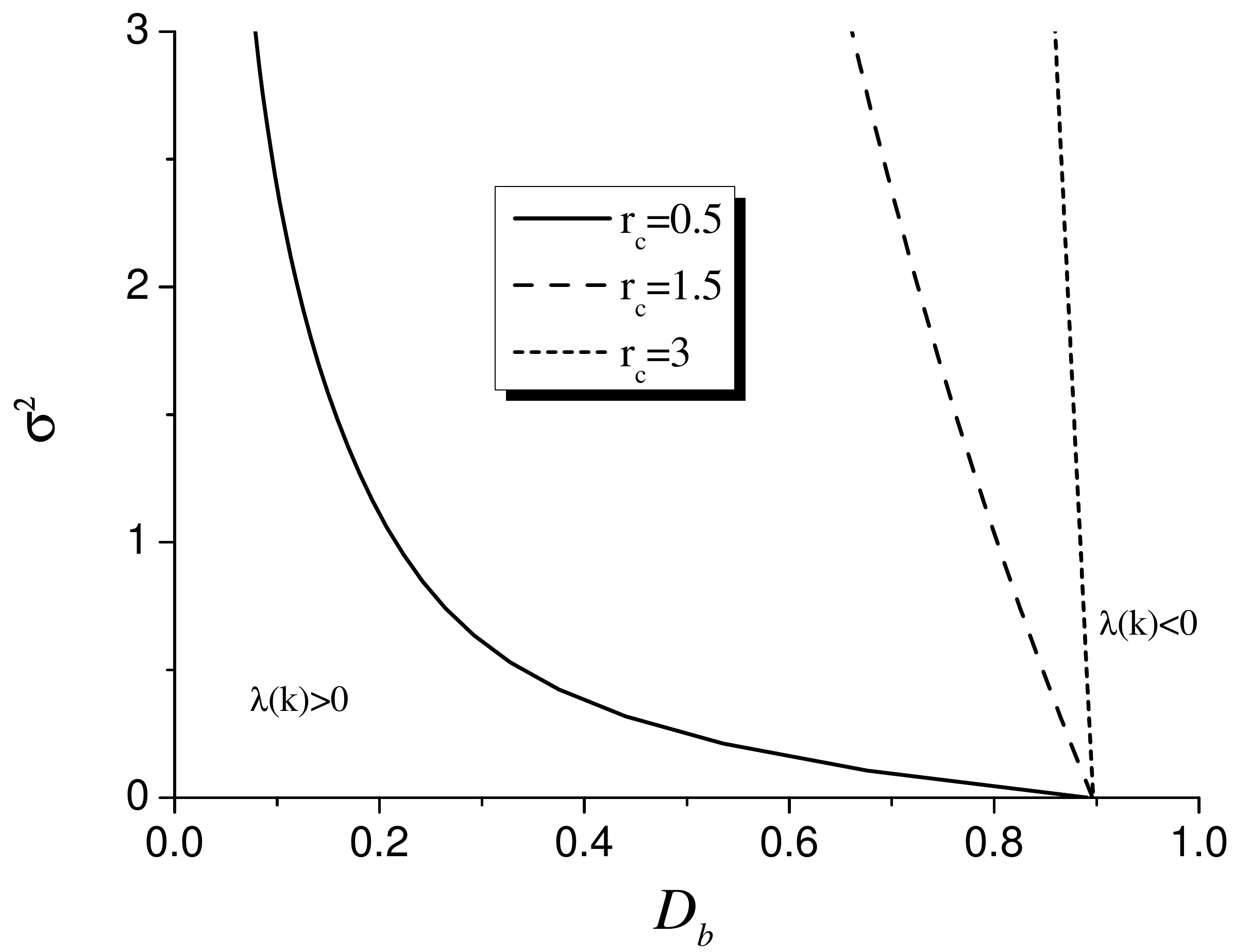}
\hspace{5mm}
\includegraphics[width=0.48\textwidth]{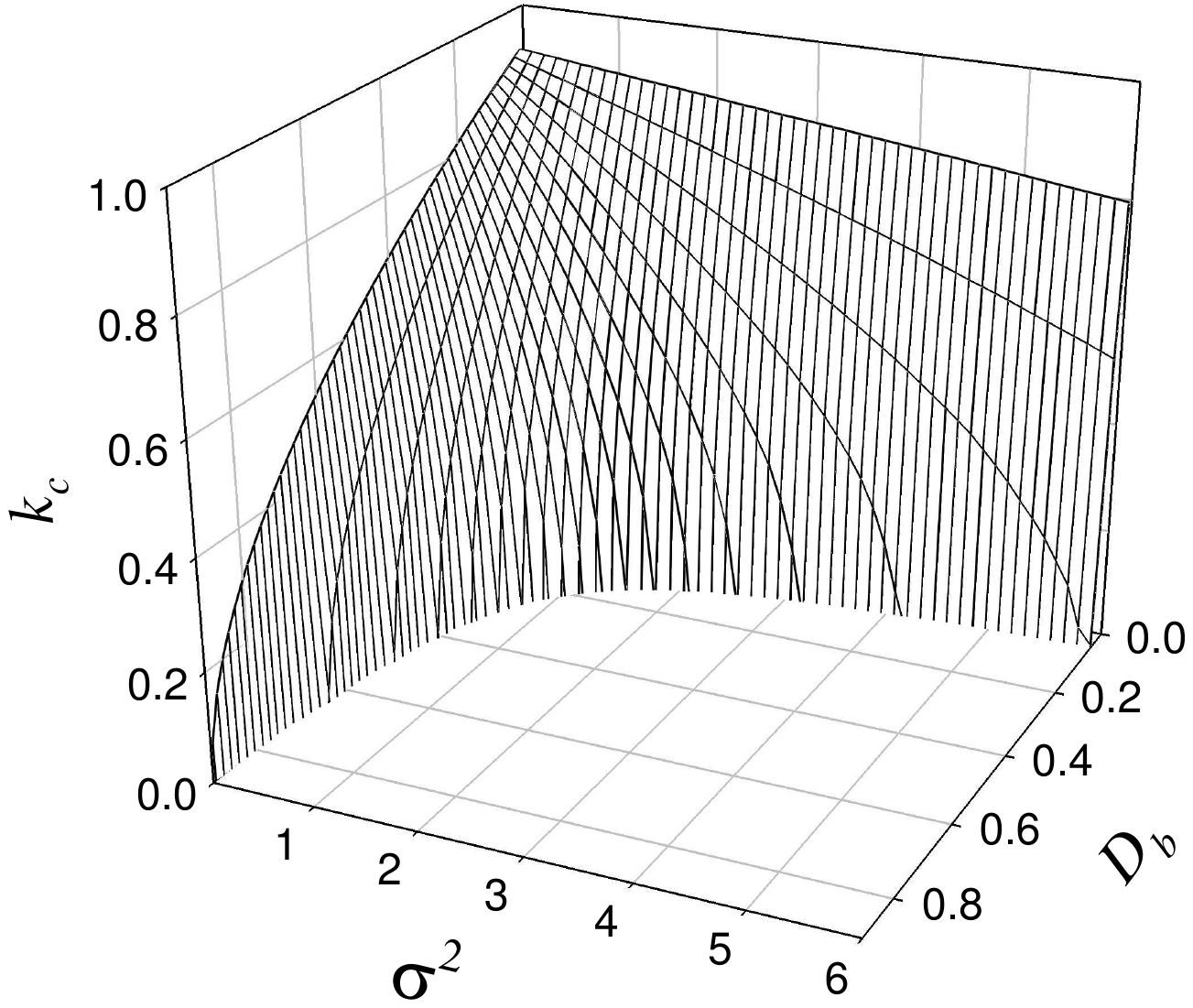}
}
\hspace{4cm} (a) \hspace{6.5cm} (b)
\caption{Stability diagram
(a) and critical wave-number dependence on $D_b$ and $\sigma^2$ (b) at $m=1$,
$\alpha=0.5$, $r_\textrm{c}=0.5$. \label{stab_phdgr}}
\end{figure}

Considering the system with two fields $\psi$ and $\phi$, let us start with
stability analysis. Averaging the system (\ref{xphi}) over noises, we get
dynamical equations for average fields in the form
\begin{equation}\label{eq20}
\begin{split}
&\partial_t \langle \psi\rangle=\nabla\cdot \langle M\nabla
\mu_\textrm{ef}\rangle-\frac{\theta}{2}\nabla\cdot \left<\nabla \partial_\psi
M\right>+\alpha\langle\phi\rangle+ \sigma^2\left[C''(\mathbf{0})\nabla^2\langle
\psi\rangle+ C(\mathbf{0})\nabla^4\langle \psi\rangle\right],\\
\frac{1}{m}&\partial_t\langle\phi\rangle=-\langle\phi\rangle-\alpha\nabla^2
 \langle \psi\rangle+e\nabla^2\langle\phi\rangle\, .
\end{split}
\end{equation}
Next, let us consider the stability of the state $(\psi=0,\phi=0)$ using Lyapunov's
analysis for fluctuations of both $\psi$ and $\phi$. A linearization of the
governing equations in the Fourier space yields
\begin{equation}
\begin{pmatrix}
\frac{{\rm d}\langle \psi\rangle}{{\rm d}t}\\[1ex] \frac{{\rm d}\langle
\phi\rangle}{{\rm d}t}
\end{pmatrix}=\begin{pmatrix}
 k^2w(k) & \alpha\\[1ex]
 m\alpha k^{2} & -m(1+ek^{2})
\end{pmatrix}\begin{pmatrix}
\langle \psi\rangle\\[1ex] \langle\phi\rangle
\end{pmatrix},
\end{equation}
where
\begin{equation}\label{w(k)}
w(k)=\alpha^2-1+\theta+D_{b}\left[1+ \sigma^2C''(0)\right]+\left[1-D_{b}\sigma^2 C(0)\right]k^2.
\end{equation}
The corresponding Lyapunov exponent takes the form
\begin{equation}\label{1.6}
2\lambda=\left[k^2w(k)-m\left(1+ek^{2}\right)\right]\pm
\sqrt{\left[k^2w(k)+m\left(1+ek^{2}\right)\right]^{2}+4m\left[1+k^2w(k)+\left(e+\alpha^2\right)k^{2}\right]}.
\end{equation}
According to the analysis of the Lyapunov exponent (\ref{1.6}), we can find
critical values for $D_b$ and $\sigma^2$, bounding the domain of unstable modes.
The corresponding stability diagram is shown in figure~\ref{stab_phdgr}~(a). It is
seen that spatial instability characterized by $\lambda(k)>0$ is possible only
inside the window for the ballistic diffusion coefficient. At the same time, a
growth in $D_b$ shrinks the interval for $\sigma^2$ where $\lambda(k)>0$. From
the stability diagram, one finds that at small $D_b$, the external noise can
sustain a spatial instability even at large intensities of fluctuations. An
increase in the correlation radius $r_\textrm{c}$ of these fluctuations enlarges
the instability domain. Therefore, strongly correlated external fluctuations are
capable of inducing spatial instability at short time scales. As
figure~\ref{stab_phdgr}~(b) shows, the critical wave-number bounding wave-number of
unstable modes decreases with $D_b$ and $\sigma^2$. Therefore, at a large
ballistic mixing intensity and the intensity of external fluctuations, long-wave
spatial instabilities should emerge over the whole system.

Let us consider the behavior of the structure function $S(k,t)$. To obtain a
dynamical equation for $S(k,t)$ in the vicinity of the point $(\psi=0,\phi=0)$,
we exploit the approach previously described by considering the system of two
equations (\ref{xphi}). Moving to the discrete representation and multiplying
every equation from the system (\ref{xphi}) by $\psi$, we arrive at the system
of two equations
\begin{equation}\label{SF2}
\begin{split}
\frac{{\rm d}S(k,t)}{{\rm d}t}=&-2k^2w(k)S(k,t)+\alpha G(k,t)+2\theta
k^2-\frac{2 k^2}{(2\pi)^2}\int{\rm d}\mathbf{q}S(q,t)
+\frac{2k^2D_{b}\sigma^2}{(2\pi)^2}\int{\rm
d}\mathbf{q}C(|\mathbf{k}-\mathbf{q}|)S(\mathbf{q},t),\\
 \frac{{\rm d}G(k,t)}{{\rm d}t}=&-2m\left[\left(1+ek^2\right)G(k,t)-\alpha k^2S(k,t)\right],
\end{split}
\end{equation}
where $G(k,t)\equiv\langle
\psi_{\mathbf{k}}(t)\phi_{-\mathbf{k}}(t)\rangle=\langle
\psi_{-\mathbf{k}}(t)\phi_{\mathbf{k}}(t)\rangle$, and $w(k)$ is given by
equation (\ref{w(k)}).

\begin{figure}[!t]
\centerline{
\includegraphics[width=0.41\textwidth]{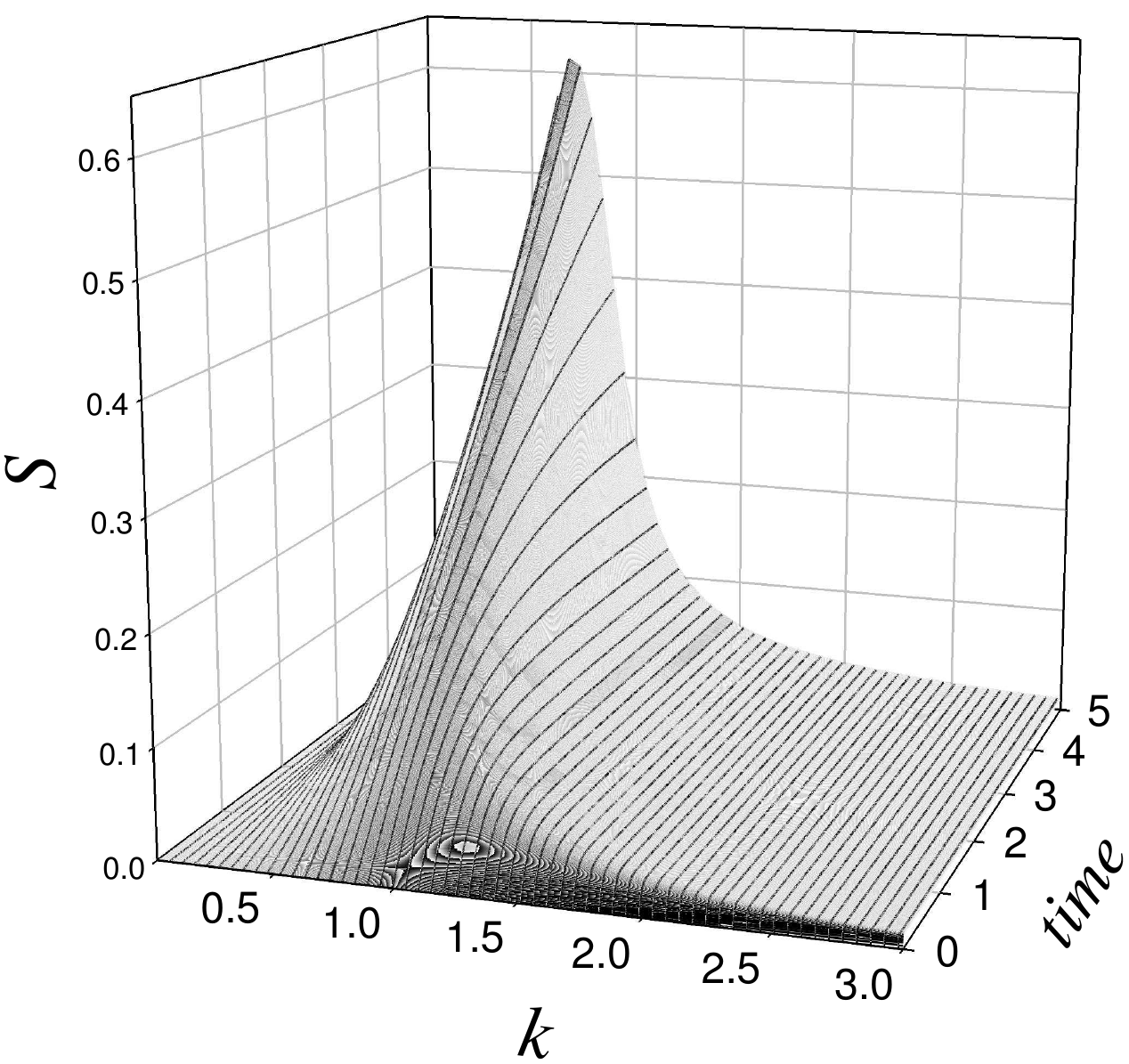}
\hspace{5mm}%
\includegraphics[width=0.51\textwidth]{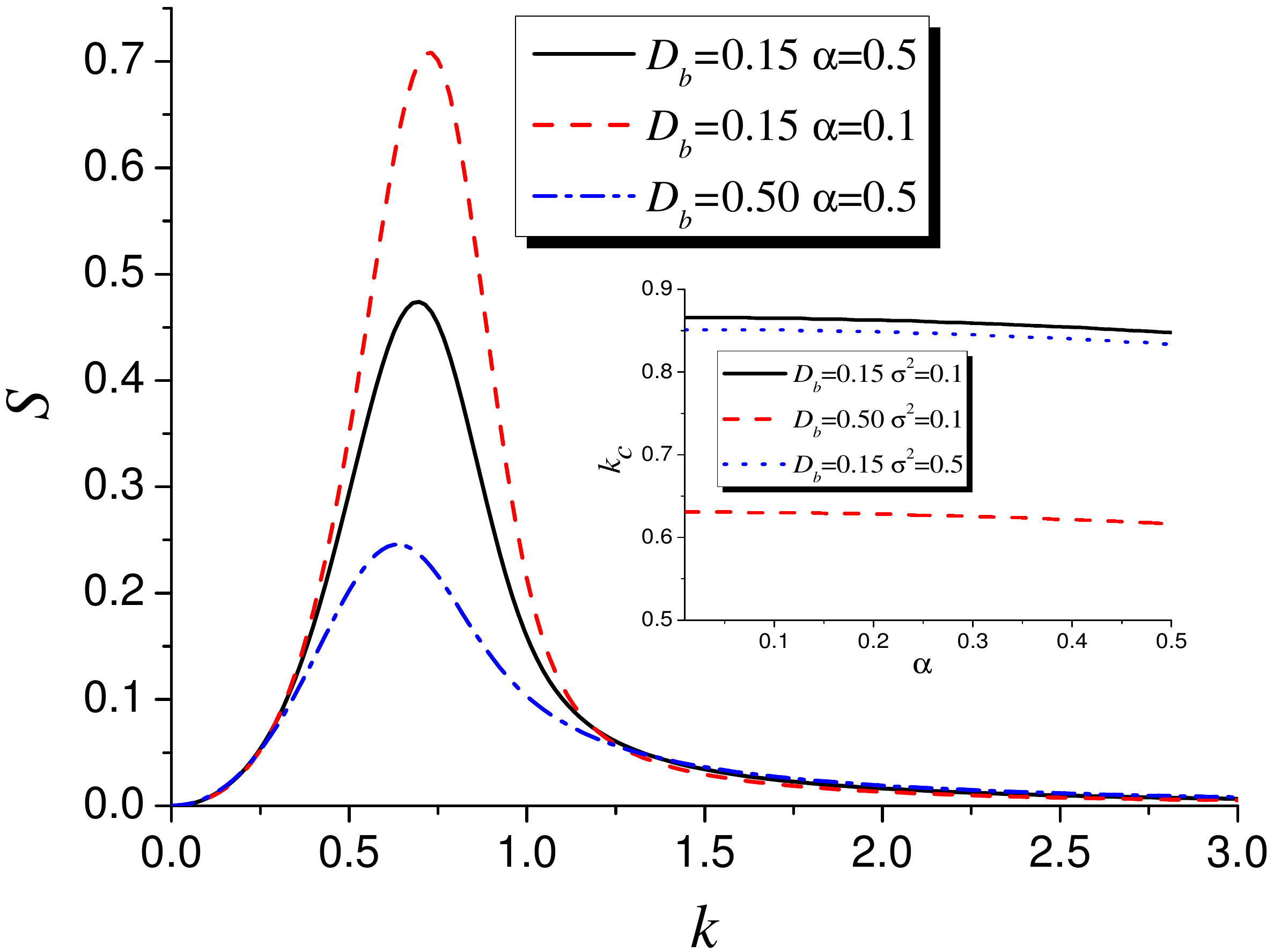}
}
\hspace{3cm} (a) \hspace{7.5cm} (b)
\caption{(Color online) The structure function
behavior for the original system: a) the dynamics of $S(k,t)$ at $D_b=0.15$,
$\sigma^2=0.1$; b) the dependence $S(k)$ at $t=10$ and a different set of the
system parameters. Other parameters are: $\theta=0.1$, $\alpha=0.5$, $e=0.2$,
$r_\textrm{c}=1.0$, $m=1.0$. \label{S(k)2}}
\end{figure}

The dynamics of $S(k,t)$ is shown in figure~\ref{S(k)2}~(a). Comparing graphs for
$S(k,t)$ related to actual and reduced models [cf.
figures~\ref{S(k)2}~(a), \ref{figS_one}~(a)], one finds that the peak of the structure
function is larger in the actual two-component dynamical model. From the
obtained dependencies for the structure function $S(k)$ shown in
figure~\ref{S(k)2}~(b) it is seen that an increase in the regular component of the
ballistic flux essentially decreases the structure function; it shifts the peak
position toward small wave-numbers. Considering the effect of the
dislocation density mechanism strength, one finds that with an increase in
$\alpha$, the wave-number of unstable modes decreases [see the insertion in
figure~\ref{S(k)2}~(b)]. At the same time, the height of the peak of the structure
function decreases at a short time scale. Therefore, the dislocation mechanism is
capable of delaying the ordering processes. Herein below, we will show that this effect can be observed by the dynamics of the average domain size.

\subsection{Numerical results}

To qualitatively describe the system behavior, we numerically solve the system
(\ref{xphi}). In simulation procedure, the Heun method was used. The system was
studied on the lattice with square symmetry of the linear size $L=128\ell$ with
periodic boundary conditions and the mesh size $\ell=0.5$; $\Delta t=0.005$ is
the time step. We take $\langle
\psi(\mathbf{0},0)\rangle=\langle \phi(\mathbf{0},0)\rangle=0$, $\langle
(\delta \psi(\mathbf{0},0))^2\rangle=\langle
(\delta\phi(\mathbf{0},0))^2\rangle=0.01$ as initial conditions. The obtained results are statistically
independent of different realizations of noise terms $\zeta$, $\xi$.

\begin{figure}[!t]
\centering
\includegraphics[width=0.98\textwidth]{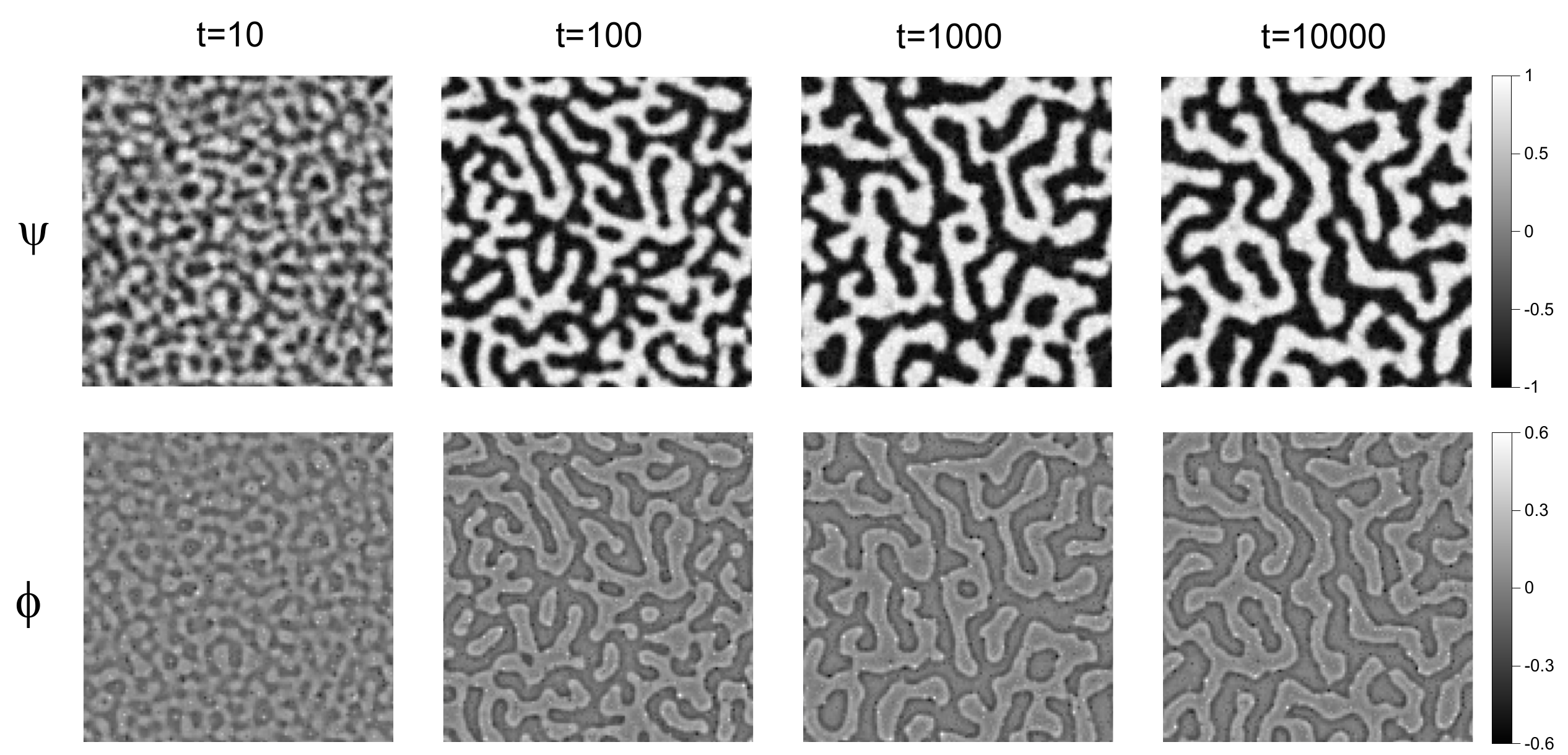}
\caption{Evolution of both the composition field $\psi$ (top panels) and
dislocation field (bottom panels) at $D_b=0.15$, $\sigma^2=0.5$, $\theta=0.1$,
$\alpha=0.5$, $e=0.2$, $r_\textrm{c}=1.0$, $m=1$.\label{evol_xphi}}
\end{figure}

Typical evolution of both the composition field $\psi$ and dislocation density
$\phi$ is shown in figure~\ref{evol_xphi}. Here, the regions of high values of both
fields $\psi$ and $\phi$ are represented by white, whereas the black areas
relate to lower values of the corresponding field. The coupling between
dislocation density and composition field is well observed. A coordinated
motion of dislocations and phase boundaries was previously observed at atomic
scale using phase field models (see references \cite{HoytHaataja,KBSG2007})

\begin{figure}[!b]
\centering
\includegraphics[width=0.75\textwidth]{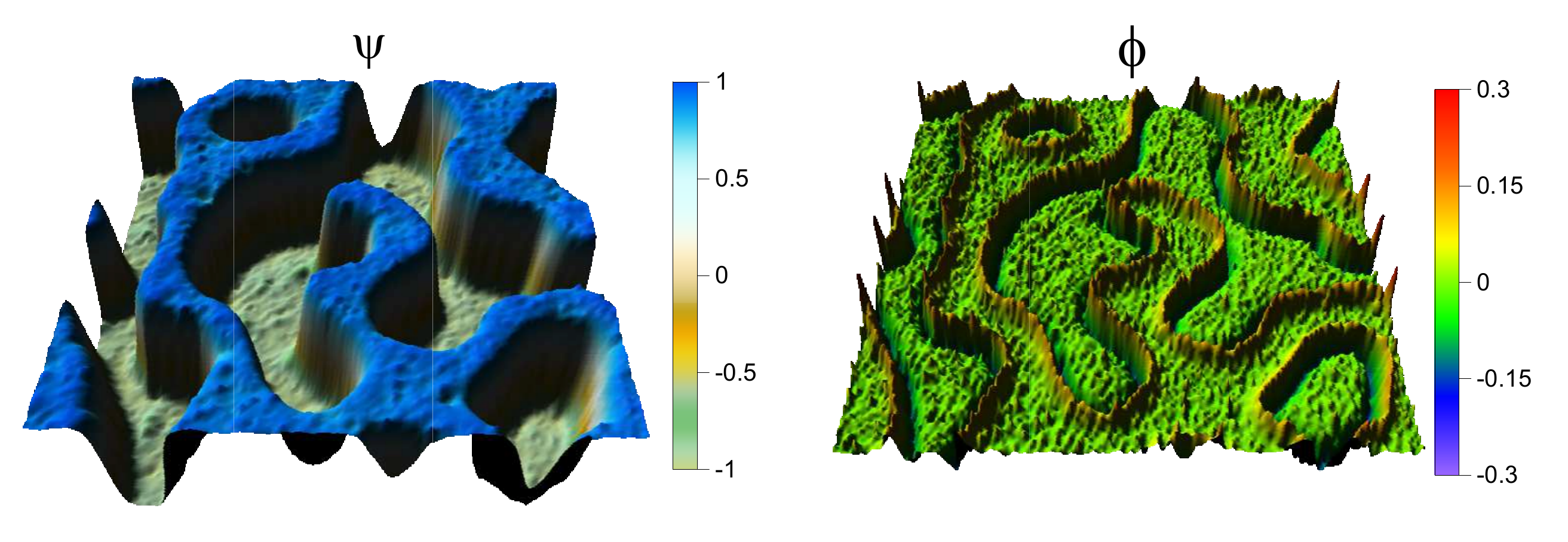}
\caption{(Color online) Snapshots of both concentration and dislocation fields at $t=10000$
shown in 3D-view, illustrating the oscillating behavior of dislocation density field
near the boundaries of two phases $\psi=+1$ and $\psi=-1$. Other parameters are:
$D_b=0.15$, $\sigma^2=0.1$, $\theta=0.01$, $\alpha=0.5$, $e=0.2$, $r_\textrm{c}=1.0$,
$m=10$.\label{dislo_struct}}
\end{figure}

It is seen that the dislocation field takes up large values in the vicinity of the
phase boundaries; inside the decomposed phases, the dislocation density is around
zero. The corresponding oscillations of $\psi$ near the interfaces indicate that
the strain energy is reduced due to the atomic size mismatch \cite{HoytHaataja}.
Therefore, misfit dislocations segregate on the boundaries. In
figure~\ref{dislo_struct}, we plot the oscillating structure of dislocation density
field $\phi$ corresponding to the distribution of the concentration field. It is
seen that in the vicinity of the boundary, $\phi$ changes the sign; inside the phases,
$\phi\simeq 0$.

To make a quantitative analysis of phase separation, let us study the dependencies
of dispersions of the fields $\psi$ and $\phi$ defined as
$J_\psi\equiv\langle(\delta \psi)^2\rangle$ and $J_\phi\equiv\langle(\delta
\phi)^2\rangle$, where $\delta \psi=\psi-\langle \psi\rangle$, $\delta
\phi=\phi-\langle \phi\rangle$. At $\langle \psi\rangle=\langle \phi\rangle=0$,
these quantities are reduced to the second statistical moments playing the role
of effective order parameters at phase decomposition processes (due to
conservation laws for $\psi$ and $\phi$). The quantity $J_\psi(t)$ is
proportional to the area below the structure function $S_k(t)$, i.e.,
$J_\psi(t)=\sum_kS_k(t)$. Therefore, the growth in $J_\psi$ (and the related
growth of $J_\phi$) corresponds to the phase decomposition process. Both moments
$J_\psi$ and $J_\phi$ grow toward nonzero stationary values. The related
stationary values $J_{\{\psi,\phi\}}^\textrm{st}=J_{\{\psi,\phi\}}(t\to \infty)$ can
be used to define two phases separated during the the long time evolution of
the system. If dislocations are excluded from the description, then one can use
only $J_\psi$ to monitor the formation of two phases. In our case, the dynamics of
phase decomposition can be described by an additional order parameter $J_\phi$
manifesting segregation of dislocations in the vicinity of the interface. From
naive consideration, one can expect that at large values of both $J_\psi$ and
$J_\phi$, one gets two well decomposed phases with an increased dislocation
density at the interface. At small $J_\psi$ and $J_\phi$, one gets a mixed state
where no large deviations in the composition field are observed. On the other
hand, it means that dislocations are distributed inside the phases.

\begin{figure}[!t]
\centering
\includegraphics[width=0.95\textwidth]{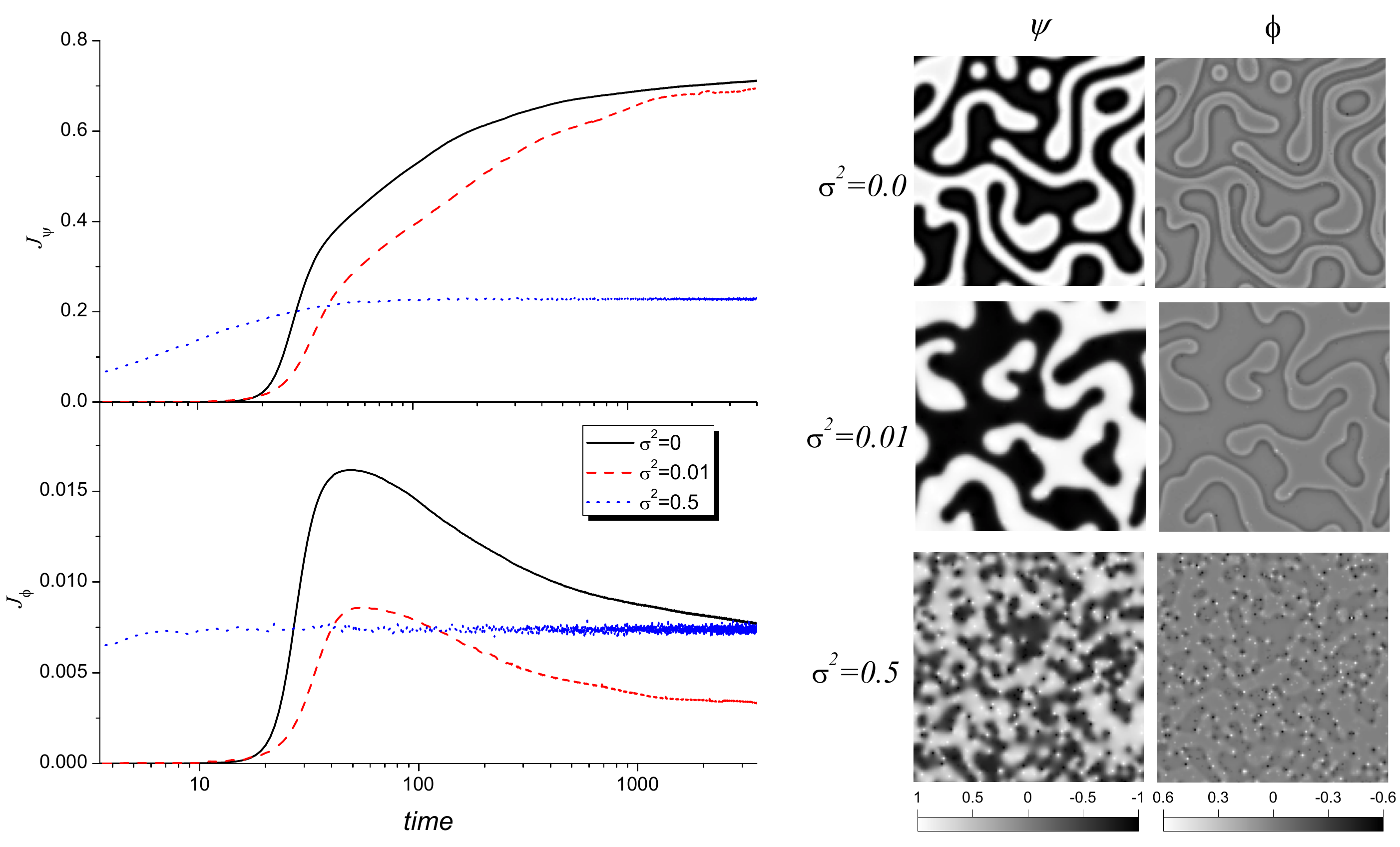}
 \caption{(Color online) The dynamics of both  order parameters $J_\psi$
 and $J_\phi$ at different values of the external noise intensity $\sigma^2$,
  and the corresponding snapshots  of both  concentration  and dislocation density fields at $t=4000$.
  Other parameters are:  $D_b=0.15$, $\theta=0.01$,
$\alpha=0.5$, $e=0.2$, $r_\textrm{c}=2.0$, $m=10$.\label{jxjfDb}}
\end{figure}

In figure~\ref{jxjfDb}, we plot the dependencies of the order parameters at different
values of the external noise intensity at other fixed system parameters. It is
seen that an increase in the noise intensity $\sigma^2$ results in small values
of both $J_\psi$ and $J_\phi$. At $\sigma^2=0$ (see solid lines in
figure~\ref{jxjfDb}), the order parameter $J_\psi$ grows toward stationary value in
a monotonous manner. This means an increase in the area under the corresponding
structure function $S(k,t)$ and the formation of well decomposed phases enriched by
atoms of different sorts (see right-hand panel illustrating the distribution of the
composition field $\psi$). The order parameter $J_\phi$ initially grows meaning
the formation of two separated phases with an increasing dislocation density in the
vicinity of the interface. At the next stage, a decaying dependence of $J_\phi$
is observed. This means an agglomeration of the domains belonging to one phase
resulting in annihilations of dislocations with opposite signs. At the late stage,
the dislocation density goes to its stationary value together with $J_\psi$. At
a small noise intensity $\sigma^2$ (see dashed curves in figure~\ref{jxjfDb}), one
observes the same dynamics of both order parameters, where $J_\psi$ and
$J_\phi$ take up low values. Here, the external noise sustains the formation of an ordered
state characterized by separated phases. This effect is caused by correlation
properties of the external noise. The deterministic part of the ballistic flux
acts in the manner opposite  to the stochastic contribution. By increasing the noise
intensity $\sigma^2$ (in the domain of a disordered phase according to the
mean-field analysis), one gets a transition toward disordered state. Here the
order parameter $J_\psi$ attains a very small stationary value (see dotted curves
in figure~\ref{jxjfDb}). The formation of a disordered state is well accompanied by time
independence of the quantity $J_\phi$. It fastly attains a small stationary value
and fluctuates around it. Here, there are no phases enriched by atoms A or B
(see the right-hand panel in figure~\ref{jxjfDb}), dislocations are distributed over the whole
system. Therefore, the effect of fluctuations characterized by large values of
$\sigma^2$ becomes larger than the correlation effects, leading to the formation of
a totally disordered state.

\begin{figure}[!t]
\centerline{
\includegraphics[width=0.52\textwidth]{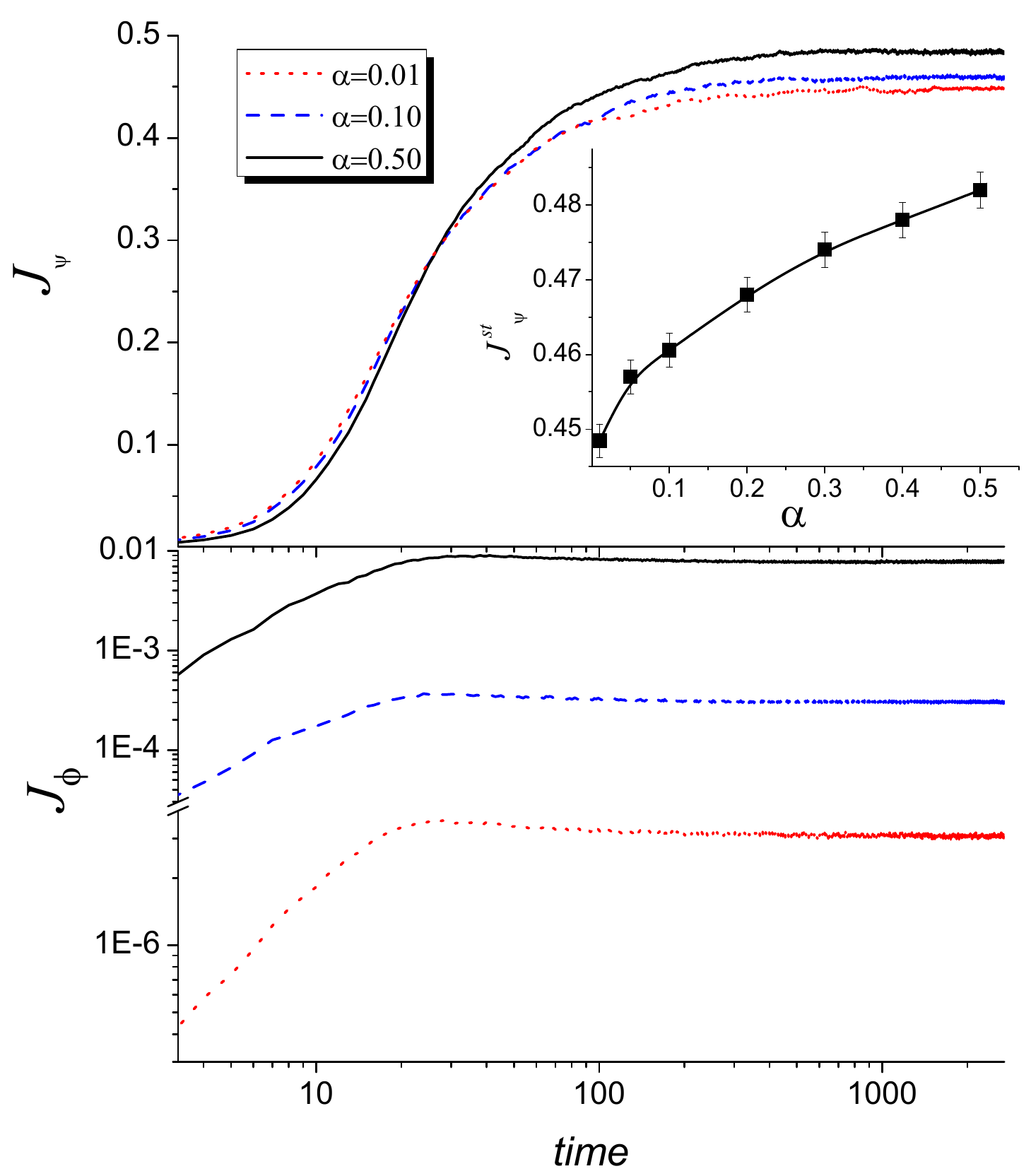}
\hspace{5mm}%
\includegraphics[width=0.41\textwidth]{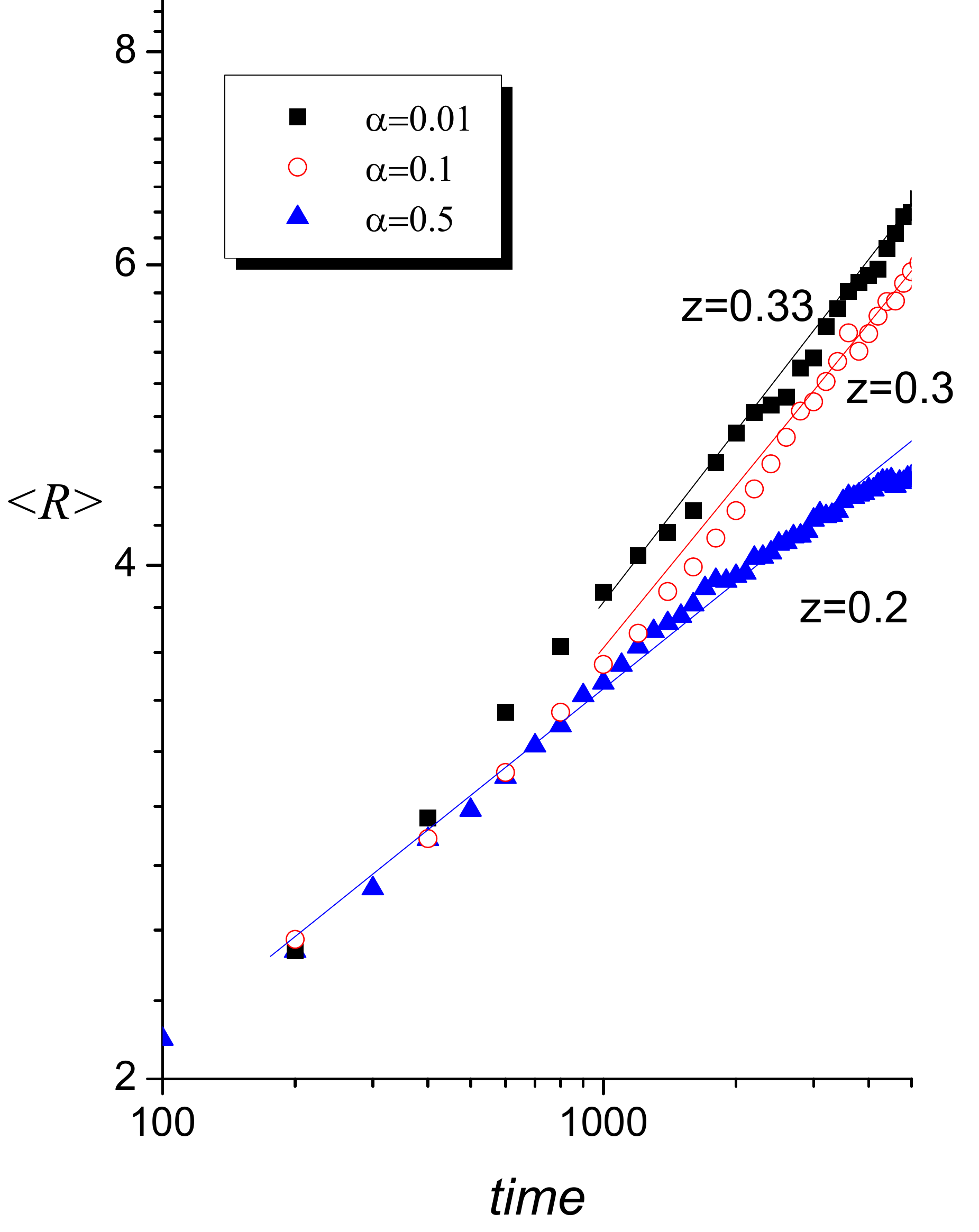}
}
 \hspace{4.5cm} (a) \hspace{6.5cm} (b)
 \caption{(Color online) (a) Dynamics of both  order parameters $J_\psi$ and $J_\phi$
 at different  $\alpha$ and  $D_b=0.12$, $\sigma^2=0.01$.
 (b) Evolution of the average domain size at different $\alpha$  and $D_b=0.1$,
 $\sigma^2=0.5$. Other parameters are:
 $\theta=0.01$, $e=0.2$, $r_\textrm{c}=2.0$, $m=10$.\label{jxjfa}}
\end{figure}

Let us consider in detail the effect of the dislocation density field onto
the dynamics of the phase decomposition. An evolution of both order parameters $J_\psi$
and $J_\phi$ at different values of the coupling constant $\alpha$ is shown in
figure~\ref{jxjfa}~(a). Here, one can see that both $J_\psi$ and $J_\phi$ increase
with $\alpha$ (the order parameter $J_\phi$ has the corresponding peak at
transition to the coarsening regime). Let us consider the behavior of the
stationary order parameter $J^\textrm{st}_\psi=J_\psi(t\to\infty)$ [see the insertion in
the upper panel in figure~\ref{jxjfa}~(a)]. It rapidly increases at small $\alpha$ and
slowly grows with $\alpha$. From the obtained results it follows that a strong
coupling between the composition field and the dislocation density field urges the
formation of the ordered state due to redistribution of dislocations over the whole
system, as well as their motion to the interfaces. Therefore, phase separation is well
sustained by a dislocation field. It is interesting to compare the dynamics of the
average domain size at different $\alpha$. According to discussions provided in
references \cite{DisloMech,DisloMech2} it is known that dislocation mechanism is capable
of changing the dynamical exponent $z$, describing the domain size growth law
$\langle R\rangle\propto t^z$. To analyze the dependence $\langle R(t)\rangle$,
we calculate the averaged value $\langle k(t)\rangle\propto1/\langle
R(t)\rangle$ according to the standard definition $\langle k(t)\rangle=\int
kS(k,t){\rm d}k/\int S(k){\rm d}k$. In the standard theory of phase
decomposition, the corresponding Lifshitz-Slyozov approach gives $z=1/3$
\cite{LS}. The same value for $z$ is observed when phase separation is
sustained by vacancy mechanism. If dislocation mechanism of phase decomposition
plays the major role, then the dynamical exponent takes up lower values $z\simeq
1/6$ \cite{DisloMech2}. In our case, we can control the strength of the dislocation
mechanism varying parameter $\alpha$. According to the results in figure~\ref{jxjfa}~(b),
one obtains $z\simeq0.33$ at $\alpha\to 0$, as Lifshitz-Slyozov theory
predicts. This result was obtained for the system subjected to a stochastic
ballistic flux with another form of the function $\tilde M(\psi)$ (see
reference \cite{EPJB2010}). It was shown that an increase in the external noise
intensity $\sigma^2$ results in disordering processes. Comparing the curves related
to $\alpha=0.1$ and $\alpha=0.5$, it follows that the dislocation mechanism delays
the dynamics of the domain sizes growth. In our case, at $\alpha=0.5$, we get a lower value
for the dynamical exponent.

\begin{figure}[!t]
\centerline{
\includegraphics[width=0.48\textwidth]{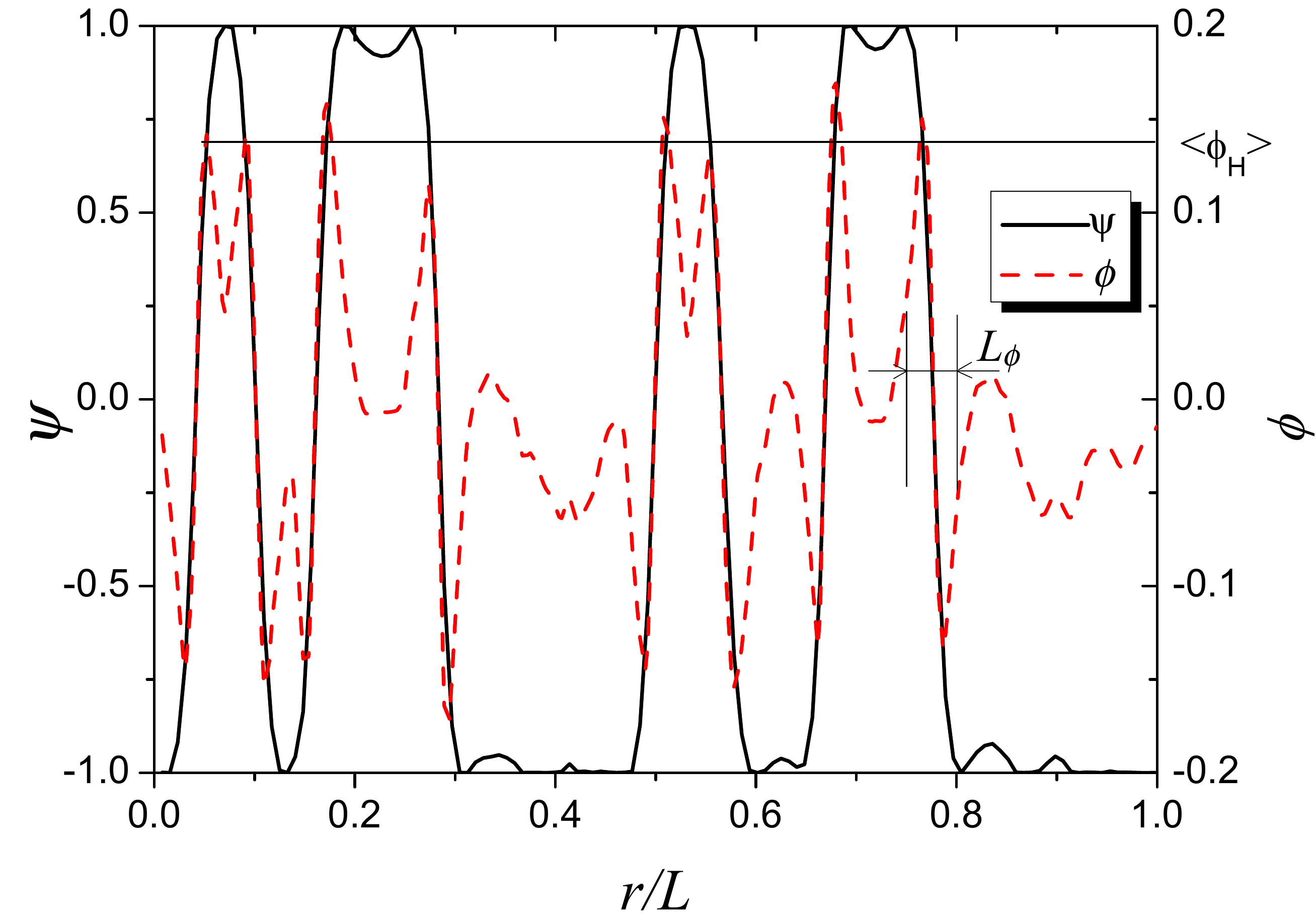}
\hspace{2mm}%
\includegraphics[width=0.5\textwidth]{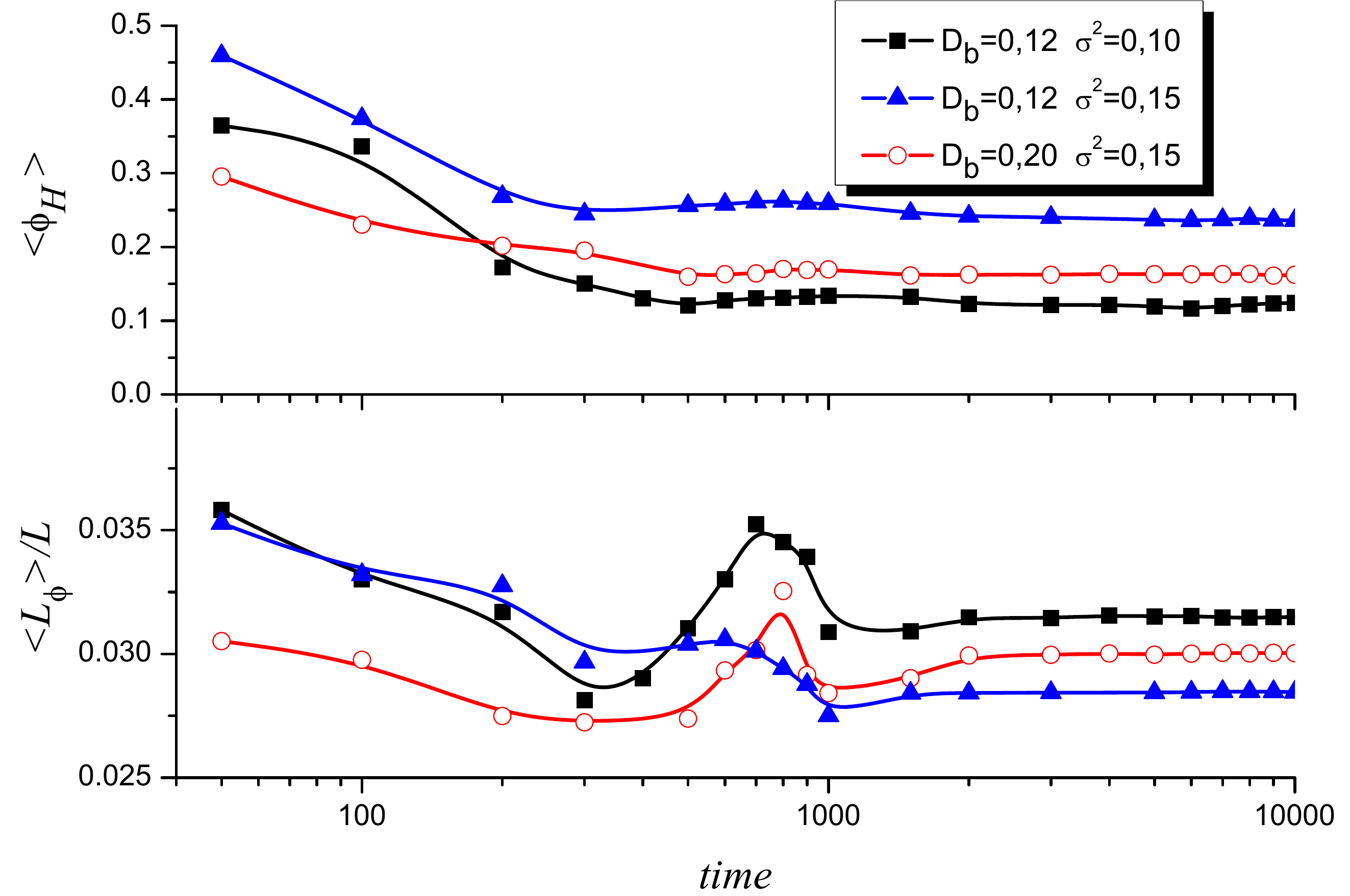}
}
\hspace{3.2cm} (a) \hspace{7.5cm} (b)
\caption{(Color online) (a) Illustration of typical one-dimensional profiles for solute
concentration $\psi$ and the field $\phi$ taken at the time $t=400$
($D_b=0.12$, $\sigma^2=0.1$). (b) Evolution of both the average maximum
dislocation density field $\langle\phi_H\rangle$ and a coherence length
$\langle L_\phi\rangle$ at different values of $D_b$ and $\sigma^2$. Other
parameters are: $\alpha=0.5$, $e=0.2$, $r_\textrm{c}=1$, $m=1$,
 $\theta=0.1$.\label{fi_HL_time}}
\end{figure}

Finally, let us consider the dynamics of the average dislocation density
$\langle\phi_H\rangle$ in the vicinity of the interfaces and the coherence
(interface) width $\langle L_\phi\rangle$, where this density decreases toward
zero value inside the separated phases.  We
calculate the quantity $\langle\phi_H\rangle$ as the mean height of $\phi(\mathbf{r})$ profile averaged over the
whole system. The coherence width $\langle L_\phi\rangle$ is calculated as the
width of the interface on the half-height of $\phi(\mathbf{r})$ profile
averaged over the system. In figure~\ref{fi_HL_time}~(a), solid and dashed lines
denote one-dimensional profiles for the composition and dislocation density
fields, respectively. The dynamics of both $\langle\phi_H\rangle$ and $\langle
L_\phi\rangle$ is shown in figure~\ref{fi_HL_time}~(b). Comparing the data related to
different sets of $D_b$ and $\sigma^2$, one finds that the dislocation density
attains a stationary value during the decomposition process. Considering
the dynamics of $\langle\phi_H\rangle$ at different external noise intensities, it
follows that the growth in $\sigma^2$ increases the values of the dislocation density.
Therefore, the external noise is capable of inducing a phase decomposition accompanied
by segregation of dislocations at interfaces. On the other hand, with an
increase in $D_b$, the quantity $\langle\phi_H\rangle$ takes up lower values. Here,
dislocations are distributed over the whole system due to homogenization of the
system produced by a regular component of the ballistic flux. The competition
between regular and stochastic components of the ballistic flux can be observed
by studying the dynamics of the averaged interface width. From the bottom panel in
figure~\ref{fi_HL_time}~(b), one finds that during the system evolution, the quantity
$\langle L_\phi\rangle$ attains a maximum. This maximum relates to the stage of
the domains growth. A coalescence regime (large domains absorb small ones) is
accompanied here by an increase in the interface width. A decrease in $\langle
L_\phi\rangle$ corresponds to a coarsening regime. At large time intervals,
$\langle L_\phi\rangle$ attains a stationary value. Comparing the corresponding
stationary values at different $D_b$, it follows that a regular component of
the ballistic flux produces a disordering accompanied by extension of the interface width
(cf. the curves marked by circles and triangles). Comparing the curves with different
$\sigma^2$, one finds that the noise is capable of playing a constructive role leading
to a decrease in the interface width. Localization of dislocations at
interfaces can be stimulated by a correlation effect of the external fluctuations
[see the curve with triangles in the upper panel in figure~\ref{fi_HL_time}~(b)]. It should
be noted that the above effect is possible only inside the domain of the ordered phase
formation. Therefore, the mentioned above constructive role of external fluctuations is
caused by their spatial correlations.

\section{Conclusions}\label{sec:5}

We have studied phase separation processes driven by the dislocation evolution
mechanism in a binary system subjected to sustained irradiation.  We describe
the irradiation effect by introducing a  ballistic flux having stochastic
properties. We assumed that these fluctuations are spatially correlated.

By taking into account different time scales of evolution of both composition
and dislocation density fields, we have initially considered a reduced model
using the adiabatic elimination procedure. In this case, the dynamics of the composition
field playing the role of a slow mode is studied. By making use of the linear stability
analysis, we have shown the constructive role of ballistic flux fluctuations.
These fluctuations induce spatial instability at a short time scale. At a large
time scale, we have used the mean-field approach allowing us to describe the properties
of phase separation. Corresponding mean-field phase diagrams illustrating
the possibility of phase decomposition are calculated. It was found that ballistic
flux components reduced to the intensity of atomic mixing (ballistic diffusion
coefficient) and the intensity of the corresponding fluctuations are capable of
controling reentrant phase separation processes. It was shown that a reentrant
character of phase separation is governed by spatial correlations of the
external noise and mobile dislocations.

Considering the dynamics of both studied fields by means of computer simulations, we
have found that the formation of domains enriched by atoms of different sort is
accompanied by an increase of the dislocation density field in the vicinity of
the interface. Inside such domains, dislocation density takes up zero values and
all dislocations characterized by different signs in two phases segregate on
interfaces. The average length of the interface decreases at the formation of the
ordered state. In the disordered state, all dislocations are distributed over
the whole system. It was shown that spatially correlated external fluctuations act
in the manner opposite to the regular component of the ballistic flux and induce
the ordered state formation. Studying the effect of the dislocation density
field onto phase decomposition we have considered the dynamics of the average
domain size at different strengths of the dislocation mechanisms. It was shown
that the universal dynamics of the average domain size delays due to a redistribution
of dislocations. We have found that at small contribution of the dislocation
density field, the corresponding universal dynamics is described by the standard
Lifshitz-Slyozov law with dynamical exponent $z=0.33$. With an increase in the
dislocation mechanism strength, this exponent takes up lower values and the domains of
new phases are characterized by smaller linear sizes.

\section*{Acknowledgements}
Fruitful discussions with Dr. V.O. Kharchenko are gratefully acknowledged.

\section*{Appendix A}\label{app:A}
\renewcommand{\theequation}{A.\arabic{equation}}

Let us represent the system on a regular $d$-dimension lattice. Within the
framework of the standard formalism of a discrete representation, the system can
be divided onto $N^d$ cells of the linear size $L=\ell N$, where $\ell$ is the
a mesh size. Then, the partial differential equation (\ref{xeq}) is reduced to
a set of usual differential equations written for every cell $i$ on a grid in
the form
\begin{equation}\label{Le_d}
\frac{{\rm d}\psi_i}{\rm{d}t}=\nabla^\textrm{L}_{ij}\tilde
M_{j}\nabla^\textrm{R}_{jl}\tilde\mu_l+ \nabla^\textrm{L}_{ij}\sqrt{\tilde
M_{j}}\xi_{j}(t)+\nabla^\textrm{L}_{ij}\nabla^\textrm{R}_{jk}\psi_k\zeta_k(t),
\end{equation}
where the index $i$ labels cells, $i=1,\ldots, N^d$; the discrete left-hand and right-hand
operators are:
\begin{equation}
\begin{split} &\nabla^\textrm{L}_{ij}=\frac{1}{\ell}(\delta_{i,j}-\delta_{i-1,j}),\qquad
\nabla^\textrm{R}_{ij}=\frac{1}{\ell}(\delta_{i+1,j}-\delta_{i,j}),\\
&\nabla^\textrm{L}_{ij}=-\nabla^\textrm{R}_{ji},\qquad\qquad\qquad
\nabla^\textrm{L}_{ij}\nabla^\textrm{R}_{jl}=\Delta_{il}=\frac{1}{\ell^2}(\delta_{i,l+1}-2\delta_{i,l}+\delta_{i,l-1}).
\end{split}
\end{equation}
Discrete correlators of stochastic sources are of the form:
\begin{equation}
 \langle\xi_{i}(t)\xi_{j}(t)\rangle=\ell^{-2}\theta\delta_{ij}\delta(t-t'),\qquad
\langle\zeta_i(t)\zeta_j(t)\rangle=D_{b}\sigma^2C_{i-j}\delta(t-t'),
\end{equation}
where $C_{i-j}$ is the discrete representation of the spatial correlation
function $C(\mathbf{r})$ which in the limit of zero correlation length becomes
$\delta_{ij}/\ell^{d}$. For the two-dimensional problem considered below, the
quantity $C_{|i-j|}$ can be computed as a discrete version of the Fourier
transform of $C(\mathbf{r}-\mathbf{r}')$ written in the form \cite{Garcia}
$C(k)=\exp\left\{-({r_\textrm{c}^2}/{2})[\sin^2(k_x/2)+\sin^2(k_y/2)]\right\}$. Noise
correlators can be calculated according to the recipes shown in
references \cite{Garcia,UJP2008,PhysA2008,EPJB2010}. Next, we consider the case $d=2$.

To calculate the internal noise correlator $\langle\sqrt{M}\xi\rangle$, we use the
Novikov theorem which can be written in the form
\begin{equation}\label{Nov1}
\left\langle \sqrt{\tilde
M_i}\xi_i(t)\right\rangle=\sum_j\int_0^t\left\langle\xi_i(t)\xi_j(t')\right\rangle
\left\langle\frac{\delta
\sqrt{\tilde M_i}}{\delta \xi_j(t')}\right\rangle{\rm
d}t'=\frac{\theta}{\ell^2}\int_0^t\left\langle\left.\frac{\delta \sqrt{\tilde
M_i}}{\delta \xi_i(t')}\right|_{t=t'}\right\rangle{\rm d}t'.
\end{equation}
Using a relation $$\left.\frac{\delta \sqrt{\tilde M_i}}{\delta
\xi_i(t')}\right|_{t=t'}=\left.\frac{{\rm d }\sqrt{\tilde M_i}}{{\rm
d}\psi_i(t)}\frac{\delta \psi_i(t)}{\delta \xi_i(t')}\right|_{t=t'}$$ and a
formal solution of the Langevin equation (\ref{Le_d}), one can write
$$\left.\frac{\delta \psi_i(t)}{\delta
\xi_i(t')}\right|_{t=t'}=\nabla^\textrm{L}_{ij}\sqrt{\tilde M_{j}}\,.$$
Substituting it
into equation (\ref{Nov1}), we get
\begin{equation}
\left\langle \sqrt{\tilde
M_i}\xi_i(t)\right\rangle=\frac{\theta}{\ell^2}\nabla^\textrm{L}_{ij}\left\langle\frac{{\rm d
}\sqrt{\tilde M_j}}{{\rm d}\psi_j}\sqrt{\tilde
M_{j}}\right\rangle=\frac{\theta}{2{\ell^2}}\nabla^\textrm{L}_{ij}\left\langle\frac{{\rm d }\tilde
M_j}{{\rm d}\psi_j}\right\rangle.
\end{equation}
Using the relation between left-hand and right-hand discrete gradient operators and moving to
the continuum limit, we get
\begin{equation}\nabla\cdot\left\langle\sqrt{\tilde
M}\xi\right\rangle=-\frac{\theta}{2}\nabla\cdot\left\langle\nabla\partial_\psi \tilde
M\right\rangle.\end{equation}

To calculate the external noise correlator
$\langle\psi\zeta\rangle$, we again use the Novikov theorem written as follows:
\begin{equation}\label{Nov2}
\langle
\psi_j\zeta_j(t')\rangle=\sum_k\int_0^t\langle\zeta_j(t)\zeta_k(t')\rangle
\left\langle\frac{\delta
\psi_j(t)}{\delta \zeta_j(t')}\right\rangle{\rm d}t'=D_b\sigma^2\sum_k C_{j-k}
\left\langle\left.\frac{\delta \psi_j(t)}{\delta \zeta_k(t')}\right|_{t=t'}\right\rangle.
\end{equation}
According to the formal solution of the Langevin equation, the corresponding
derivative with respect to $\zeta$ takes the form
$\left<\delta\psi_j/\delta\zeta_k|_{t=t'}\right>=\Delta_{jk}\left<\psi_k\right>$.
Next, substituting it into equation (\ref{Nov2}) and using a discrete representation of
the Laplacian, we find the sum over the index $k$ allowing us to write
\begin{equation}
\langle
\psi_j\zeta_j\rangle=\frac{D_b\sigma^2}{\ell^{2}}\left[C_1\langle\psi_{j+1}\rangle+C_{-1}\langle\psi_{j-1}\rangle-2C_0\langle\psi_{j}\rangle\right].
\end{equation}
Acting by Laplacian operator onto this construction, we finally obtain in
continuum limit
\begin{equation}
\Delta\langle
\psi\zeta\rangle=D_b\sigma^2
\left[C(\mathbf{0})\nabla^4\langle\psi\rangle+C''(\mathbf{0})\Delta\langle\psi\rangle\right].
\end{equation}

\section*{Appendix B}\label{app:B}
\renewcommand{\theequation}{B.\arabic{equation}}

To obtain a dynamical equation for the structure function, we need to construct
a dynamical equation for the two-point correlation function $\langle
\psi_i\psi_j\rangle$. In our computation procedure, we use the procedure described
in reference \cite{LSA}. Multiplying the linearized Langevin equation (\ref{Le_d})
written for $\psi_j$ onto $\psi_i$ and doing the same procedure for the
linearized Langevin equation written for $\psi_i$, we add these two equations.
This procedure allows us to obtain a dynamical equation for $\langle
\psi_i\psi_j\rangle$ written in the form
\begin{eqnarray}\label{cor_psi0}
\frac{{\rm d}\langle \psi_i\psi_j\rangle}{{\rm d}t}
& = &
\Delta_{ik}\left[\left(D_{b}-1+\alpha^2\right)\langle \psi_j\psi_k\rangle-\left(1-\alpha^2e\right)\Delta_{kl} \langle \psi_l\psi_j\rangle +
	\langle \psi_j\psi_k\zeta_k\rangle\right] \nonumber\\
& + & \langle \psi_i\nabla^\textrm{L}_{jk}\xi_{k}\rangle
	-2\langle \psi_i\nabla^\textrm{L}_{jk}\psi_k\xi_{k}\rangle\nonumber\\
& + &
\Delta_{jk}\left[\left(D_{b}-1+\alpha^2\right)\langle \psi_i\psi_k\rangle-\left(1-\alpha^2e\right)\Delta_{kl} \langle \psi_l\psi_i\rangle +
	\langle \psi_i\psi_k\zeta_k\rangle\right]
\nonumber\\
& + & \langle \psi_j\nabla^\textrm{L}_{ik}\xi_{k}\rangle -2\langle
 \psi_j\nabla^\textrm{L}_{ik}\psi_k\xi_{k}\rangle,
\end{eqnarray}
where the sum runs over the repeating indexes. Using the Novikov theorem with
recipe shown in Appendix A, one can calculate the following correlators:
\begin{equation}\label{cor_psi1}
\begin{split}
&\langle
\psi_i\nabla^\textrm{L}_{jk}\xi_{k}\rangle=-\frac{\theta}{\ell^2}\Delta_{ij}\,,
\qquad
\langle
\psi_i\nabla^\textrm{L}_{jk}\psi_k\xi_{k}\rangle=-\frac{\theta}{2\ell^2}\left(\Delta_{jk}\langle
\psi_k\psi_j\rangle +\Delta_{ik}\langle \psi_k\psi_j\rangle\right),
\\
 &\langle \psi_j\psi_k\zeta_k\rangle=D_{b}\sigma^2\left(\Delta_{kl}C_{k-l}\langle
\psi_l\psi_j\rangle+\Delta_{jl}C_{k-l}\langle \psi_l\psi_k\rangle\right).
\end{split}
\end{equation}
Introducing the structure function $S(\mathbf{k},t)\equiv \langle
\psi_{\mathbf{k}}(t)\psi_{-\mathbf{k}}(t)\rangle$ in the discrete space
$$S_\nu(t)=(N\ell)^{-2}\langle \psi_\nu(t)\psi_{-\nu}(t)\rangle$$ with Fourier
components
$$\psi_\nu(t)=\ell^2\sum_m \re^{-\ri\mathbf{r}_m\mathbf{k}_\nu}\psi_m(t),\qquad
\psi_m(t)=(N\ell)^{-2}\sum_\nu \re^{\ri\mathbf{r}_m\mathbf{k}_\nu}\psi_\nu(t)$$ one
can define the derivative $$\frac{{\rm d}S_\nu}{{\rm d}t}=(\ell/N)^2
\re^{\ri\mathbf{k}_\nu(\mathbf{r}_j-\mathbf{r}_m)}\frac{{\rm d}\langle
\psi_m\psi_j\rangle}{{\rm d}t}.$$

In the following computations, one needs to exploit definitions:
\begin{equation}
\left(\frac{\ell}{N}\right)^2\re^{\ri\mathbf{k}_\nu(\mathbf{r}_j-\mathbf{r}_i)}\Delta_{is}\langle\psi_s\psi_j\rangle
=\widehat{D}_\nu S_\nu(t),
\end{equation}
\begin{equation}
\Delta_{is}\langle\psi_s\psi_j\rangle=(N\ell)^{-4}\re^{\ri\mathbf{k}_\nu\mathbf{r}_j}
\re^{\ri\mathbf{k}_\mu\mathbf{r}_i}\widehat{D}_\mu\langle\psi_\mu\psi_\nu\rangle,
\end{equation}
\begin{equation}
2\left(\frac{\ell}{N}\right)^2\re^{\ri\mathbf{k}_\nu(\mathbf{r}_j-\mathbf{r}_i)}C_{|i-j|}\langle\psi_i\psi_j\rangle=(N\ell)^{-2}\sum_\mu
C_{\mu}S_{\mu-\nu}(t),
\end{equation}
where $\widehat{D}_\nu
=\ell^{-2}\big(\sum_{|i|=1}\cos(\mathbf{k}_\nu\mathbf{r}_i)-1\big)$ can be
understood as the Fourier transform of the discrete Laplacian. Using the above
definitions and equation (\ref{cor_psi0}) with correlators (\ref{cor_psi1}), we arrive
at the equation for the quantity $S_\nu(t)$:
\begin{eqnarray}\label{strf_discr}
\frac{{\rm d}S_\nu(t)}{{\rm
d}t}&=&2\widehat{D}_\nu\left[\alpha^2-1+\theta+D_{b}+
D_b\sigma^2\Delta_{0m}C_m-\widehat{D}_\nu\left(1-D_b\sigma^2C_1\right)\right]S_\nu(t)\nonumber\\
&-&2\theta
\widehat{D}_\nu+2\widehat{D}_\nu(N\ell)^{-2}\sum_\mu S_\mu(t)
-2D_{b}\sigma^2(N\ell)^{-2}\widehat{D}_\nu \sum_\mu C_{\nu-\mu}S_\mu(t).
\end{eqnarray}
In continuous and thermodynamical limit ($N\to\infty$, $\ell\to 0$), we obtain
a dynamical equation for the structure function in the form
\begin{eqnarray}\label{F2_d}
\frac{{\rm d}S(k,t)}{{\rm d}t}&=&-2k^2\left\{\alpha^2-1+\theta+D_{b}(1+
\sigma^2C''(0))+k^2\left[1-\alpha^2e -D_{b}\sigma^2 C(0)\right]\right\}S(k,t) \nonumber \\
&+& 2\theta
k^2-\frac{2 k^2}{(2\pi)^2}\int{\rm d}\mathbf{q}S(q,t)
+\frac{2k^2D_{b}\sigma^2}{(2\pi)^2}\int{\rm
d}\mathbf{q}C(|\mathbf{k}-\mathbf{q}|)S(\mathbf{q},t).
\end{eqnarray}%

\section*{Appendix C}\label{app:C}
\renewcommand{\theequation}{C.\arabic{equation}}

To obtain the probability density distribution $\mathcal{P}([\psi],t)$ in
$d$-dimensional space, let us start with definitions:
\begin{equation}
\mathcal{P}([\psi],t)=\left\langle\prod\limits_{i=1}^{N^d}\rho_i(t)\right\rangle
\equiv\left\langle\rho(t)\right\rangle,\quad
\rho_i(t)=\overline{\delta(\psi_i(t)-\psi_i)}_\textrm{IC}\,,
\end{equation}
where $\ldots_\textrm{IC}$ and $\langle\ldots \rangle$ are the averages over initial
conditions and fluctuations, respectively. To obtain the corresponding
Fokker-Planck equation, we use a standard technique and exploit the stochastic
Liouville equation for the distribution $\rho(t)$ in the form
\begin{equation}
\partial_t\rho=-\frac{\partial}{\partial \psi_i}(\dot \psi_i \rho).
\end{equation}
Inserting the expression for the time derivative from equation (\ref{Le_d}) and
averaging over the noise, we get
\begin{equation}
\partial_t \mathcal{P}=-\frac{\partial}{\partial \psi_i}\left(\nabla^\textrm{L}_{ij}M_{j}\nabla^\textrm{R}_{jl}\tilde\mu_l \right)\mathcal{P}
-\frac{\partial}{\partial \psi_i}\left( \langle
\nabla^\textrm{L}_{ij}\sqrt{M_j}\xi_{j}(t)\rho\rangle + \langle\Delta_{ij}
\psi_{j}\zeta_j(t)\rho\rangle\right).
\end{equation}

Correlators in the second term can be calculated by means of the Novikov
theorem, that at $\ell=1$ gives~\cite{Novikov}
\begin{equation}
\begin{split}
&\left\langle
\nabla^\textrm{L}_{ij}\sqrt{M_j(t)}\xi_{j}(t)\rho\right\rangle
=\theta\int\limits_0^t{\rm
d}t'\delta_{jk}\delta(t-t')\left\langle\frac{\delta
\nabla^\textrm{L}_{ij}\sqrt{M_j(t)}\rho}{\delta \xi_{k}(t') }\right\rangle,\\
&\qquad\langle\Delta_{ij} \psi_{j}\zeta_j(t)\rho\rangle =D_{b}\sigma^2\int\limits_0^t{\rm
d}t'C_{|j-k|}\delta(t-t')\left\langle\frac{\delta \Delta_{ij} \psi_{j}\rho}{\delta
\zeta_k(t') }\right\rangle.
\end{split}
\end{equation}
Introducing notations $g_{ij}=\{(\nabla_L)_{ij}\sqrt{M_j},\Delta_{ij}
\psi_{j}\}$, $\lambda=\{\xi,\zeta\}$, for the last multiplier, one has
\begin{equation}
\frac{\delta g_{ij}\rho(t)}{\delta \lambda_k(t') }=-\sum_l
g_{ij}\frac{\partial}{\partial \psi_k}\frac{\delta \psi_l(t)}{\delta
\lambda_k(t')}\bigg|_{t=t'}\rho.
\end{equation}
Using a formal solution of the Langevin equation, the response functions take
up the form
\begin{equation}
\left.\frac{\delta \psi_l(t)}{\delta
\xi_{k}(t')}\right|_{t=t'}=\nabla^\textrm{L}_{lk}\sqrt{M_k},\quad \left.\frac{\delta
\psi_l(t)}{\delta \zeta_k(t')}\right|_{t=t'}=\Delta_{lk}\psi_{k}\,.
\end{equation}

After some algebra, we obtain the Fokker-Planck equation for the total
distribution $\mathcal{P}$ in the discrete space
\begin{equation}\label{dFPE}
\partial_t\mathcal{P}=-\frac{\partial}{\partial \psi_i}\left(\nabla^\textrm{L}_{ij}M_{j}\nabla^\textrm{R}_{jl}\tilde\mu_l
 \right)\mathcal{P}-\theta\frac{\partial}{\partial
\psi_i}\nabla^\textrm{L}_{ij}\sqrt{M_j}\frac{\partial}{\partial
\psi_j}\nabla^\textrm{R}_{ji}\sqrt{M_i}\mathcal{P}+D_{b}\sigma^2\frac{\partial}{\partial
\psi_i}\Delta_{ij} \psi_j\frac{\partial}{\partial \psi_l}C_{|j-k|}\Delta_{kl}
\psi_l \mathcal{P},
\end{equation}
where the relations between left-hand and right-hand gradient operators are used.

To proceed, let us obtain an evolution equation for the single-site probability
distribution
$$P_i(t)=\int\Bigg[\prod_{k\ne i}{\rm d
}\psi_k\Bigg]\mathcal{P}.$$ After integration one has
\begin{equation}
\frac{\partial P_i(t)}{\partial t}=\frac{\partial}{\partial
\psi_i}\Delta_{ij}\langle \mathcal{M}_j\rangle P_i(t),
\end{equation}
where
\begin{eqnarray}
 \mathcal{M}_j&=&M_j\left[-\frac{\partial f}{\partial
\psi_j}-\frac{\alpha^2+D_{b} }{M_j}
\psi_j+\frac{1}{2d}\left(1-\frac{\alpha^2e}{M_j}\right)\Delta_{jr}\psi_r\right]\nonumber\\
&-&\theta
\sqrt{M_j}\frac{\partial}{\partial \psi_j}\sqrt{M_j}+D_{b}\sigma^2 \psi_{j}
\frac{\partial}{\partial \psi_n}\Delta_{mn} C_{|j-n|}\psi_n.
\end{eqnarray}

In the stationary case with no flux, one arrives at the equation
\begin{equation}
\Delta_{ij}\langle \mathcal{M}_j\rangle P_{s}(\psi_i)=0,
\end{equation}
where $P_\textrm{s}$ is a stationary distribution function. By taking $i=j$, dropping the
subscripts, we arrive at the mean-field stationary equation \cite{IGTS99}
\begin{eqnarray}\label{st_FPE}
-hP_\textrm{s} &=& \left\{ M
  \left[ -\partial_{\psi}f-\frac{\alpha^2+D_{b}}{M} \psi+\left(1-\frac{\alpha^2e}{M}\right)(\eta-\psi)\right]\right.\nonumber\\
   &-&\left.\frac{\theta}{2}
  \partial_\psi M+2dD_{b}\sigma^2 \psi \left[C_1\eta\frac{\partial}{\partial \psi
}-C_0\frac{\partial}{\partial \psi}\psi\right]\right\}P_\textrm{s}\,.
\end{eqnarray}
Here we have used the mean-field approximation, allowing us to write
\begin{equation}
\Delta_{ij}\psi_j\equiv\left( \sum\limits_{nn(i)}\psi_{nn(i)}-2d
\psi_i\right)\rightarrow 2d (\langle \psi\rangle-\psi), \qquad \eta\equiv \langle
\psi\rangle,
\end{equation}
where $nn(i)$ denotes the nearest neighbors of a given site. The mean-field value
$\eta$ and the integration constant $h$ should be defined self-consistently.
Equation (\ref{st_FPE}) has the solution in the form
\begin{equation}\label{pdf_X}
P_\textrm{s}(\psi,\eta,h)=N\exp\left(\int^\psi {\rm
d}\psi'\frac{\Theta(\psi',\eta,h)}{\Xi(\psi';\eta)}\right),
\end{equation}
where
\begin{equation}\label{pdf_X1}
\begin{split}
\Theta(\psi,\eta,h)&=-M
\partial_{\psi}f-\left(\alpha^2+D_{b}\right)\psi+ \left(M-\alpha^2e\right)(\eta-\psi)-\frac{\theta}{2}
\partial_{\psi}M-2d D_{b}\sigma^2C_0\psi+h,\\
\Xi(\psi;\eta)&= \theta M+2d D_{b}\sigma^2\psi(C_0\psi-C_1\eta)
\end{split}
\end{equation}
the normalization constant is a function of the mean-field $\eta$ and the
constant $h$:
\begin{equation}\label{pdf_X11}
N\equiv N(\eta,h)=\left[\int\limits_{-1}^{1}{\rm d}\psi\exp\left(\int^\psi {\rm
d}\psi'\frac{\Theta(\psi',\eta,h)}{\Xi(\psi';\eta)}\right)\right]^{-1}.
\end{equation}

\newpage

\ukrainianpart

\title{Дослідження процесів фазового розшарування за наявності дислокацій в бінарних системах, підданих опроміненню }%
\author{ Д.О. Харченко, О.М. Щокотова, А.І. Баштова. І.О. Лисенко}
\address{Інститут прикладної фізики НАН України,  вул. Петропавлівська 58, 40000 Суми, Україна}
\makeukrtitle
\begin{abstract}
\tolerance=3000%
Проведено дослідження процесів фазового розшарування за дислокаційним
механізмом в бінарних системах, підданих дії опромінення. Опромінення
описується атермічним перемішуванням атомів, за рахунок уведення балістичного
потоку, що має просторово-скорельовану стохастичну складову. При вивченні
динаміки росту доменів показано, що дислокаційний механізм уповільнює процес
упорядкування. Встановлено, що просторові кореляції шуму балістичного потоку
стимулюють сегрегацію ядер дислокацій в околі міжфазних границь, ефективно
зменшуючи ширину міжфазного шару. Розглянуто конкуренцію між регулярною та
стохастичною компонентами балістичного потоку.

\keywords фазове розшарування, опромінення, шум

\end{abstract}


\begin{thebibliography}{99}

\bibitem{Shulson} Schulson E.M., J. Nucl. Mater., 1979, \textbf{83}, 239; \doi{10.1016/0022-3115(79)90610-X}.

\bibitem{Russel} Russel K.C., Prog. Mater. Sci., 1984, \textbf{28}, No.~3--4, 229; \doi{10.1016/0079-6425(84)90001-X}.

\bibitem{OM1} Bernas H.,  Attane J.P.,  Heinig K.H.,  Halley D.,  Ravelosona D.,
Marty A., Auric P., Chappert C.,  Samson Y., Phys. Rev. Lett., 2003, \textbf{91}, 077203; \doi{10.1103/PhysRevLett.91.077203}.

\bibitem{OM2} Wei L.,  Lee Y.S.,  Averback R.S.,  Flynn C.P., Phys. Rev. Lett., 2000, \textbf{84}, 6046; \doi{10.1103/PhysRevLett.84.6046}.

\bibitem{OM3} Lee Y.S.,  Flynn C.P.,  Averback R.S., Phys. Rev. B, 1999, \textbf{60}, 881; \doi{10.1103/PhysRevB.60.881}.

\bibitem{JohnsonOrlov} Johnson R.A., Orlov A.N., {Physics of Radiation Effects in Crystals}, Elsevier, Amsterdam, 1986.

\bibitem{EPJB2012} Kharchenko V.O., Kharchenko D.O., Eur. Phys. J. B,
2012, \textbf{85}, 383; \doi{10.1140/epjb/e2012-30522-3}.

\bibitem{UJP2013} Kharchenko D.O., Kharchenko V.O., Bashtova A.I., Ukr. J. Phys.,  2013, \textbf{58}, No.~10, 993.

\bibitem{CMPh2013} Kharchenko V.O., Kharchenko D.O., Condens. Matter Phys., 2013, \textbf{16}, 33001; \doi{10.5488/CMP.16.33001}.

\bibitem{REDS2014} Kharchenko D.O., Kharchenko V.O., Bashtova A.I., Radiat. Eff. Defects Solids,  2014, \textbf{169}, No.~5, 418; \\ \doi{10.1080/10420150.2014.905577}.

\bibitem{Siegel}  Siegel S., Phys. Rev., 1949, \textbf{75}, 1823; \doi{10.1103/PhysRev.75.1823}.

\bibitem{Martin} Martin G.,  Phys. Rev. B, 1984, \textbf{30}, 1424; \doi{10.1103/PhysRevB.30.1424}.

\bibitem{VaksKamyshenko} Vaks V.G., Kamyshenko V.V., Phys. Lett. A, 1993, \textbf{177}, 269; \doi{10.1016/0375-9601(93)90039-3}.

\bibitem{Abromeit96} Matsumara S., Tanaka Y., M\"uller S., Abromeit C.,
J. Nucl. Mater., 1996, \textbf{239}, 42; \\ \doi{10.1016/S0022-3115(96)00431-X}.

\bibitem{MartinBellon97} Martin G., Bellon P., In: Solid State Physics: Advances in Research and Applications, vol.~50, Ehrenreich~H.,
  Spaepen F. (Eds.), Academic Press,  New York, 1997,  pp.~189--331; \doi{10.1016/S0081-1947(08)60605-0}.

\bibitem{Was}Was G.S., {Fundamentals of Radiation Material Science}, Springer-Verlag, Berlin, 2007, pp.~433--490.

\bibitem{Wagner}Wagner W., Poerschke R., Wollemberger H., J. Phys. F, 1982, \textbf{12}, 405; \doi{10.1088/0305-4608/12/3/008}.

\bibitem{Garner} Garner F.A., McCerthy J.M., Russell K.C., Hoyt J.J., J. Nucl. Mater.,
1993, \textbf{205}, 411; \\ \doi{10.1016/0022-3115(93)90105-8}.

\bibitem{Nak89} Nakai K., Kinoshita C., J. Nucl. Mater.,  1989, \textbf{169}, 116; \doi{10.1016/0022-3115(89)90526-6}.

\bibitem{Nak91_1}Nakai K., Kinoshita C., Nishimura N.,  J. Nucl. Mater., 1991, \textbf{179--181}, 1046; \doi{10.1016/0022-3115(91)90271-8}.

\bibitem{Nak91_2}Asai Y., Isobe Y., Nakai K., Kinoshita C., Shinohara K.,  J. Nucl. Mater., 1991, \textbf{179--181}, 1050; \\ \doi{10.1016/0022-3115(91)90272-9}.

\bibitem{Stag1}Wilkers P., J. Nucl. Mater., 1979, \textbf{83}, 166; \doi{10.1016/0022-3115(79)90602-0}.

\bibitem{Stag2}Frost H.J., Russell K.C., Acta Metall., 1982, \textbf{30}, 953; \doi{10.1016/0001-6160(82)90202-4}.

\bibitem{Stag3}Abromeit C., Naundorf V., Wollenberger H., J. Nucl. Mater.,
1988, \textbf{155--157}, 1174; \\ \doi{10.1016/0022-3115(88)90491-6}.

\bibitem{Haa02}Haataja M., Muller J., Rutenberg A.D., Grant M., Phys. Rev. B, 2002,
\textbf{65}, 165414; \doi{10.1103/PhysRevB.65.165414}.

\bibitem{Haa04}Haataja M., Leonard F., Phys. Rev. B, 2004, \textbf{69}, 081201; \doi{10.1103/PhysRevB.69.081201}.

\bibitem{Haa05}Haataja M., Mahon J., Provatas N., Leonard F., Appl. Phys. Lett.,
2005, \textbf{87}, 251901; \doi{10.1063/1.2147732}.

\bibitem{HoytHaataja}Hoyt J.J., Haataja M., Phys. Rev. B, 2011, \textbf{83}, 174106; \doi{10.1103/PhysRevB.83.174106}.

\bibitem{EnriqueBellon9902} Enrique R., Bellon P., Phys. Rev. B, 1999, \textbf{60}, 14649; \doi{10.1103/PhysRevB.60.14649}.

\bibitem{EnriqueBellon00}Enrique R., Bellon P., Phys. Rev. Lett., 2000, \textbf{84}, 2885; \doi{10.1103/PhysRevLett.84.2885}.

\bibitem{EnriqueBellon02}  Liu J.-W., Bellon P., Phys. Rev. B, 2002, \textbf{66}, 020303(R); \doi{10.1103/PhysRevB.66.020303}.

\bibitem{EnriqueBellon04}Enrique R., Bellon P., Phys. Rev. B, 2004, \textbf{70}, 224106; \doi{10.1103/PhysRevB.70.224106}.

\bibitem{Yanovski} Dubinko V.I., Tur A.V., Yanovsky V.V., Radiat. Eff. Defects Solids, 1990, \textbf{112}, 233; \doi{10.1080/10420159008213049}.

\bibitem{EPJB2010}Kharchenko D.O., Lysenko I.O., Kokhan S.V.,  Eur. Phys. J. B, 2010, \textbf{76}, 37; \doi{10.1140/epjb/e2010-00172-8}.

\bibitem{UJP2010} Kharchenko D.O., Lysenko I.O., Kharchenko V.O., Ukr. J. Phys., 2010, \textbf{55}, No.~11, 1225.

\bibitem{PhysA2010}Kharchenko D., Lysenko I., Kharchenko V.,  Physica A, 2010, \textbf{389}, 3356; \doi{10.1016/j.physa.2010.04.027}.

\bibitem{CEJP2011} Kharchenko D., Kharchenko V., Lysenko I., Cent. Eur. J. Phys., 2011, \textbf{9}, No.~3, 698; \doi{10.2478/s11534-010-0076-y}.

\bibitem{DisloMech}Vengrenovich R.D., Moskalyuk A.V., Yarema S.V., Ukr. J. Phys., 2006, \textbf{51}, No.~3, 307.

\bibitem{DisloMech2}Vengrenovich R.D., Moskalyuk A.V., Yarema S.V., Phys. Solid State, 2007, \textbf{49}, No.~1, 11; \doi{10.1134/S1063783407010039}
[Fiz. Tverd. Tela, 2007, \textbf{49}, No.~1, 13 (in Russian)].

\bibitem{CahnHill}Cahn J.W., Hilliard J.E., J. Chem. Phys., 1958, \textbf{28}, 258; \doi{10.1063/1.1744102}.

\bibitem{Martin90} Martin G., Phys. Rev. B, 1990, \textbf{41}, 2279; \doi{10.1103/PhysRevB.41.2279}.

\bibitem{Cahn1961} Cahn J.W., Acta Metall., 1961, \textbf{9}, 795; \doi{10.1016/0001-6160(61)90182-1}.

\bibitem{Hoyt} Hoyt J.J., {Phase Transitions}, McMaster Innovation Press, Hamilton, 2010.

\bibitem{32}Nelson D.R., Phys. Rev. B, 1978, \textbf{18}, 2318; \doi{10.1103/PhysRevB.18.2318}.

\bibitem{33}Nelson D.R., Halperin B.I., Phys. Rev. B, 1979, \textbf{19}, 2457; \doi{10.1103/PhysRevB.19.2457}.

\bibitem{34}Bak\'o B., Hoeffelner W., Phys. Rev. B, 2007, \textbf{76}, 214108; \doi{10.1103/PhysRevB.76.214108}.

\bibitem{35}Gromma I., Bak\'o B., Phys. Rev. Lett., 2000, \textbf{84}, 1487; \doi{10.1103/PhysRevLett.84.1487}.

\bibitem{36}Enomoto Y., Iwata S., Surf. Coat. Technol., 2003, \textbf{169--170}, 233; \doi{10.1016/S0257-8972(03)00088-4}.

\bibitem{Halpering} Hohenberg P.C., Halperin B.I., Rev. Mod. Phys., 1977, \textbf{49},
435; \doi{10.1103/RevModPhys.49.435}.

\bibitem{AbrMartin99} Abromeit C., Martin G., J. Nucl. Mater., 1999, \textbf{271--272}, 251; \doi{10.1016/S0022-3115(98)00712-0}.

\bibitem{AppPhysLett2005}Leonard F., Haataja M., Appl. Phys. Lett., 2005,
\textbf{86}, 181909; \doi{10.1063/1.1922578}.

\bibitem{Novikov} Novikov E.A., Sov. Phys. JETP, 1965, \textbf{20}, 1290.

\bibitem{Garcia} Garcia-Ojalvo J., Sancho J.M., {Noise in Spatially
Extended Systems}, Springer-Verlag, New York, 1999.

\bibitem{UJP2008}  Kharchenko D.O., Dvornichenko A.V., Lysenko I.O., Ukr. J. Phys.,
2008, \textbf{53}, No.~9, 917.

\bibitem{IGTS99} Ibanes M., Garcia-Ojalvo J., Toral R., Sancho J.M., Phys. Rev. E, 1999,
\textbf{60}, 3597; \\ \doi{10.1103/PhysRevE.60.3597}.

\bibitem{GO93} Garcia-Ojalvo J., Lacasta A.M., Sancho J.M.,  Toral R.,
Europhys. Lett., 1998, \textbf{42}, 125; \doi{10.1209/epl/i1998-00217-9}.


\bibitem{GO2001} Ibanes M., Garcia-Ojalvo J., Toral R.,
Sancho J.M., Phys. Rev. Lett., 2001, \textbf{87}, 020601; \\ \doi{10.1103/PhysRevLett.87.020601}.

\bibitem{PhysA2008} Kharchenko D.O., Dvornichenko A.V., Physica A, 2008, \textbf{387}, 5342; \doi{10.1016/j.physa.2008.05.041}.

\bibitem{GMS83} Gunton J.D., Miguel M.S., Sahni P.S., In: {Phase Transtions and Critical
Phenomena}, Domb~C., Lebowitz~J.L. (Eds.), Academic Press, New York, 1983.

\bibitem{KBSG2007}Elder K.R., Berry J., Stefanovic P., Grant M., Phys. Rev. B, 2007, \textbf{75}, 064107; \doi{10.1103/PhysRevB.75.064107}.

\bibitem{LS} Lifshitz I.M.,  Slyozov V.V., J. Phys. Chem. Solids, 1961,  \textbf{19}, 35; \doi{10.1016/0022-3697(61)90054-3}.

\bibitem{LSA} Ibanes M., Garcia-Ojalvo J., Toral R., Sancho J.M., In:
Stochastic Processes in Physics, Chemistry, and Biology,
Lecture Notes in Physics Series, vol.~557, Freund~J., P\"oschel~T. (Eds.), Springer, Berlin, 2000, pp.~247-256; \\ \doi{10.1007/3-540-45396-2_23}.


\end{thebibliography}
\end{document}